**Kinetic Study of the Gas-Phase Reaction between Atomic Carbon and Acetone. Low Temperature Rate Constants and Hydrogen Atom Product Yields**

Kevin M. Hickson,[1,*] Jean-Christophe Loison,[1] and Valentine Wakelam[2]

[1]Institut des Sciences Moléculaires ISM, CNRS UMR 5255, Univ. Bordeaux, 351 Cours de la Libération, F-33400, Talence, France

[2]Laboratoire d'astrophysique de Bordeaux, CNRS, Univ. Bordeaux, B18N, allée Geoffroy Saint-Hilaire, F-33615 Pessac, France

**Abstract**

The reactions of ground state atomic carbon, C($^3$P), are likely to be important in astrochemistry due to the high abundance levels of these atoms in the dense interstellar medium. Here we present a study of the gas-phase reaction between C($^3$P) and acetone, CH$_3$COCH$_3$. Experimentally, rate constants were measured for this process over the 50-296 K range using a continuous-flow supersonic reactor, while secondary measurements of H($^2$S) atom formation were also performed over the 75-296 K range to elucidate the preferred product channels. C($^3$P) atoms were generated by In-situ pulsed photolysis of carbon tetrabromide, while both C($^3$P) and H($^2$S) atoms were detected by pulsed laser induced fluorescence. Theoretically, quantum chemical calculations were performed to obtain the various complexes, adducts and transition states involved in the C($^3$P) + CH$_3$COCH$_3$ reaction over the $^3$A'' potential energy surface, allowing us to better understand the reaction pathways and help to interpret the experimental results. The derived rate constants are large, (2-3) × 10$^{-10}$ cm$^3$ s$^{-1}$, displaying only weak temperature variations; a result that is consistent with the barrierless nature of the reaction. As this reaction is not present in current astrochemical networks, its influence on simulated interstellar acetone abundances is tested using a gas-grain dense interstellar cloud model. For interstellar modelling purposes, the use of a temperature independent value for the rate constant, $k_{\text{C}+\text{CH}_3\text{COCH}_3} = 2.2 \times 10^{-10}$ cm$^3$ s$^{-1}$, is recommended. The C($^3$P) + CH$_3$COCH$_3$ reaction decreases gas-phase CH$_3$COCH$_3$ abundances by as much as two orders of magnitude at early and intermediate cloud ages.

# 1 Introduction

Carbon bearing compounds are ubiquitous throughout the Universe, with carbon being present as the fourth most abundant element following H, He and O. In interstellar space, elemental carbon is considered to be mostly present as $C^+$ in diffuse clouds (here elements with an ionization potential < 13.6 eV are generally considered to be fully ionized), before transforming into neutral atomic carbon C($^3$P) and initially, simple molecular species (CO, CH, $CH_2$, $CH_3$) in denser regions of interstellar space. Observations of neutral atomic carbon towards dense molecular clouds such as TMC-1,[1] OMC-1[2,3] and Barnard 5[4] through the C($^3P_1$-$^3P_0$) and C($^3P_2$-$^3P_1$) fine structure transitions in the submillimeter-wave range at 492.162 GHz and 809.345 GHz clearly indicate that neutral atomic carbon and CO coexist deep inside these clouds. Indeed, recent models[5] predict that neutral atomic carbon abundances could be as high as $10^{-4}$ (relative to $H_2$) during the early stages of cloud evolution. In this respect, it is important to investigate the reactivity of C($^3$P) with a wide range of interstellar molecules to evaluate their potential importance to the overall chemistry of these regions. Although there are numerous studies of the kinetics[6-9] and dynamics[7, 9-13] of C($^3$P) reactions with a wide range of unsaturated hydrocarbons, there are relatively few studies of its reactions with organic molecules containing other types of functional groups such as those found in complex organic molecules (COMs). Over the last few years, we have begun to address this issue by examining the reactivity of C($^3$P) with important COMs such as $CH_3OH$[14] and $CH_3CN$[15] and other molecules that are thought to be formed primarily on interstellar ices such as $N_2O$.[5]

At the present time, the chemical processes leading to the formation of many of the observed interstellar COMs are not well constrained in a general sense. Among them, acetone, $CH_3COCH_3$, was first detected in the dense molecular cloud Sgr B2 (OH) by Combes et al.[16] in 1987 with a derived column density of $5 \times 10^{13}$ cm$^{-2}$ corresponding to a relative abundance (with respect to $H_2$) of $5 \times 10^{-11}$. Its detection was subsequently confirmed by Snyder et al. [17] who observed several transitions towards Sgr B2 (OH), Sgr B2 (M) and Sgr B2 (N), while confirming that acetone emission arose predominantly from the hot molecular core Sgr B2 (N-LMH). They derived much higher relative abundances, $(4\text{-}30) \times 10^{-10}$, compared to those obtained by Combes et al. [16] Very recently, acetone has been detected at the cyanopolyyne peak position of the prototypical cold dark cloud TMC-1, through the QUIJOTE survey observations with an abundance of $1.4 \times 10^{-11}$.[18] Acetone has also been observed in several other interstellar environments including the high mass star forming regions GAL 31.41+0.31, GAL 034.3+00.2, and GAL 10.47+00.03[19] with column densities in the range $(0.1\text{-}1.3) \times 10^{15}$ cm$^{-2}$ and towards the low-mass protostar IRAS 16293-2422[20] with a column density of $1.7 \times 10^{16}$ cm$^{-2}$.

In terms of its production in the gas-phase ISM, acetone can be formed from acetaldehyde ($CH_3CHO$) by the radiative association reaction $CH_3^+ + CH_3CHO \rightarrow (CH_3)_2CHO^+ + h\nu$ followed by dissociative recombination $(CH_3)_2CHO^+ + e^- \rightarrow CH_3COCH_3 + H$, but certain models[21] have shown that this process largely underestimates observed abundances. The neutral-neutral reaction between atomic oxygen and $C_3H_7$ is also considered to be a minor source.[22] Acetone is lost through ion-molecule reactions with several cations such as $H_3^+$, $HCO^+$, $He^+$ and $H_3O^+$.[22] Although it does not react with the other abundant interstellar atomic radicals N($^4$S) and O($^3$P) at low temperature in the gas-phase, acetone has been shown to react with small hydrogen bearing radicals such as CH[23] and OH.[24,25] Although the OH + acetone reaction has not been conclusively demonstrated to lead to reaction products through direct measurements (as opposed to complex stabilization through collisions with the carrier gas), this process is nonetheless included in current astrochemical databases, while rate constants for the CH + acetone reaction have never been measured so this reaction is not currently included. [22]

Instead, modeling studies have highlighted the probable influence of the formation of complex organic molecules (COMs) in general through neutral-neutral reactions on the surface of ice-covered interstellar dust grains. It has been suggested that the reaction of $CH_3$ with $CH_3CO$[26] followed by chemical desorption of a fraction of the surface formed products could be an important source of gas-phase acetone. Nevertheless, as the observed gas-phase acetone abundances are not well reproduced by astrochemical models, this seems to indicate that important gas-phase or surface reactions leading to acetone formation are poorly constrained or missing entirely from current networks.

In order to improve the current knowledge of acetone chemistry in the interstellar medium, we have undertaken an experimental investigation of the gas-phase $C(^3P)$ + acetone reaction over the 50-296 K temperature range using a continuous supersonic flow apparatus, coupled with pulsed laser photolysis and pulsed laser induced fluorescence for $C(^3P)$ atom formation and detection respectively. In addition to the kinetic measurements, we have also performed measurements of the product channels leading to atomic hydrogen formation at room temperature and below. In conjunction with new electronic structure calculations of the various complexes, intermediates and transition states over the triplet potential energy surface (PES), this work has allowed us to identify the major product channels for interstellar modeling studies. Finally, we have tested the effect of the $C(^3P)$ + acetone reaction on a dense cloud model while updating the chemistry leading to acetone formation and loss in both the gas-phase and on interstellar ices.

The experimental and theoretical methods employed are described in sections 2 and 3 respectively, while the results of this work are presented in section 4. The astrochemical simulations and the implications of this study for interstellar acetone and related species are described in section 5, followed by our conclusions in section 6.

## 2 Experimental Methods

All the experiments described here were performed using a supersonic flow reactor, also known by the French acronym CRESU (Cinétique de Réaction en Écoulement Supersonique Uniforme or Reaction Kinetics in a Uniform Supersonic Flow). This apparatus, which has been described in detail in earlier work, [27,28] employs axisymmetric Laval nozzles to generate low temperature flows with uniform density, temperature and velocity characteristics over a known distance from the nozzle, allowing the kinetics of fast gas-phase reactions to be investigated (with rate constants > $10^{-12}$ cm$^3$ s$^{-1}$). Since these early studies, numerous modifications have been made, notably allowing the detection of atomic species in ground (C($^3$P), [29, 30] H($^2$S) [31,32]) and excited (O($^1$D), [33,34] N($^2$D) [35,36]) electronic states through their transitions in the vacuum ultraviolet wavelength range. Measurements were performed at five different temperatures during the present work (296 K, 177 K, 127 K, 75 K, 50 K) with N$_2$ or Ar as the carrier gases. Three different Laval nozzles were used to access temperatures below 296 K (one nozzle was employed with both N$_2$ and Ar, allowing flow temperatures of 177 K and 127 K to be generated respectively). The nozzle flow characteristics are summarized in Table 2 of Hickson et al.[31] For each temperature, the distance between the Laval nozzle exit and the observation axis was set to the maximum value for which the flow conditions could still be considered optimal, providing us with the maximum possible time to record kinetic profiles. These characteristic distances were derived in earlier experiments measuring the supersonic flow impact pressure as a function of distance from the nozzle using a Pitot tube.

The nozzle was removed to perform room temperature experiments, with a reduced flow velocity to eliminate pressure gradients within the reactor.

Acetone was introduced into the flow via a temperature- and pressure-controlled bubbler system upstream of the Laval nozzle. Here, a small fraction of the carrier gas flow (< 25 sccm) was diverted into a bubbler containing acetone maintained at room temperature and a pressure around 400 Torr. The acetone laden carrier gas was then flowed into a cold trap maintained at 17 °C, allowing the gas-phase acetone concentration at the trap exit to be calculated precisely using its saturated vapour pressure at this temperature.[37] The output of the cold trap was connected to the Laval nozzle reservoir through a tube heated to 80 °C to prevent condensation before further dilution by the remaining carrier gas flows. In this case, the use of the heated tube was simply a precautionary measure as no condensation losses were ever observed during experiments with and without heating.

In common with our earlier studies of $C(^3P)$ reactivity at low temperature, carbon tetrabromide ($CBr_4$) was used as the precursor molecule in the present work. These molecules were entrained in the flow by passing a small fraction (40 – 70 sccm) of the total flow over solid $CBr_4$ maintained at a known fixed pressure and room temperature in a separate vacuum flask. $CBr_4$ concentrations in the supersonic flow were estimated to be lower than $2.6 \times 10^{13}$ cm$^{-3}$ based on its saturated vapour pressure. $C(^3P)$ atoms were produced directly in the cold supersonic flow by the 266 nm pulsed laser photolysis of $CBr_4$ at 10 Hz. This beam (with pulse energies of 30-35 mJ for a 5 mm beam diameter) was steered in through the back of the reactor via a fused silica window oriented at the Brewster angle to reduce window fluorescence, and was coaligned along the supersonic flow, exiting by the nozzle throat and a second Brewster angled window positioned at the back of the reservoir. In this way, a column of $C(^3P)$ atoms was produced along the entire length of the supersonic flow, with the nascent

C($^3$P) concentration being identical at any axial position, due to the weak attenuation of the UV beam by CBr$_4$ ($\sigma_{\text{CBr}_4}(266 \text{ nm}) = 1 \times 10^{-18}$ cm$^2$). It should be noted that under our experimental conditions, CBr$_4$ photolysis at 266 nm also produces some excited state C($^1$D) atoms at the level of 10-15 %.[14] The possible interferences brought about by the presence of these atoms are discussed in section 4.

C($^3$P) atoms were detected directly during the present kinetic experiments by pulsed laser induced fluorescence in the vacuum ultraviolet wavelength range through the 2s$^2$2p$^2$ $^3$P$_2$ → 2s$^2$ 2p5d $^3$D$_3$° transition at 115.803 nm. Tunable radiation around this wavelength was produced by the second harmonic generation of the output from a narrowband dye laser at 695 nm, itself pumped by the second harmonic output (532 nm) of a 10 Hz pulsed Nd:YAG laser. The residual 695 nm radiation after doubling was eliminated from the UV beam at 347 nm using two dichroic mirrors coated for maximal reflection at 355 nm. The pulse energy was typically 8-9 mJ for a 5 mm beam diameter. Part of the discarded visible radiation was fed into the input coupler of a wavelength meter allowing us to track the probe laser wavelength variations in real time. The 347 nm beam itself was steered using right angled prisms and was focused into a gas cell attached to the reactor at the level of the observation axis (and at right angles to the supersonic flow) via a 75 cm arm. The cell contained 50 Torr of xenon with 160 Torr of argon for optimal phase matching, allowing us to generate the required tunable VUV radiation around 115.8 nm by frequency tripling. For the H-atom product detection experiments, tunable radiation around the Lyman-α transition at 121.567 nm was generated in a similar manner, with the dye laser operating at 729 nm and a mixture of 210 Torr of krypton and 540 Torr of argon in the tripling cell.

As the VUV beam was divergent as it left the cell, a MgF$_2$ lens was used instead of a plane window at the cell exit to recollimate the beam before entering the reactor where it was allowed to interact with the supersonic flow. Due to the difference in refractive index of $MgF_2$ at VUV and UV wavelengths, the UV beam diverged along the input arm, allowing a large fraction of this radiation to be blocked by a series of circular diaphragms positioned at various distances between the cell and the reactor, although a small fraction of the UV beam did still interact with the supersonic flow. As the arm was open to the reactor, containing species such as $CBr_4$ and $CH_3COCH_3$ which absorb strongly in the VUV range, this zone was flushed continuously by a flow of Ar or $N_2$ during the experiments. To check that the residual UV beam entering the reactor did not affect the measurements, several test experiments were performed. First, at the start of the initial experiments measuring rate constants and H-atom branching ratios, the cell was evacuated to ensure that no spurious signals were observed in the absence of VUV light. Similarly, preliminary experiments were also performed in the absence of the photolysis laser (with and without gas in the tripling cell) to check for the formation of $C(^3P)$ and/or $H(^2S)$ by either the VUV beam or the residual UV beam. No signal from $C(^3P)$ was ever observed, but a small H-atom fluorescence signal was observed when $CH_3COCH_3$ was present in the flow and gas was present in the tripling cell, due to $CH_3COCH_3$ photolysis by the VUV probe beam at 121.567 nm. In this case, this signal did not vary as a function of time (as the probe laser energy was fixed). Additionally, a larger H-atom fluorescence signal was observed with $CH_3COCH_3$ in the flow with both the photolysis and VUV probe lasers on, arising from $CH_3COCH_3$ photolysis at 266 nm. An explanation of how we disentangled the contribution of the H-atom photolysis signal from the H-atom reactive signal is provided in section 4.3. The fluorescence emission from C- or H- atoms within the flow was detected at right angles to both the supersonic flow and the VUV beam. The detection system consisted of a solar blind photomultiplier tube (PMT), isolated from the reactor by a LiF window, while a series of circular diaphragms were placed along the detection axis to reduce the detection of scattered light. Within the isolated region, which was evacuated by an oil-free vacuum pump to prevent atmospheric absorption losses, a LiF lens focused the VUV emission onto the PMT photocathode. The output signal from the PMT was fed into a gated boxcar integrator, with the boxcar system, oscilloscope (for signal visualization and gate positioning) and both lasers being synchronized by a digital pulse generator. C($^3$P) and H($^2$S) fluorescence signals were recorded as a function of time, with each time point consisting of 30 individual laser shots. For each temporal scan, at least 100 time points were typically recorded, with at least 15 of them at negative time delays (that is, with the probe laser firing prior to the photolysis laser). These points allowed us to establish the baseline level, which consisted of any scattered light signals and, in the case of the H-atom detection experiments, the small contribution of H-atom fluorescence from $CH_3COCH_3$ photolysis by the VUV probe laser as described above.

The carrier gas flows used in the present experiments Ar (Messer 99.999 %)and $N_2$ (Messer 99.999 %) were regulated by calibrated mass-flow controllers. All gases including the rare gases used in the tripling cell, Xe (Linde 99.999 %) and Kr (Linde 99.999 %), were flowed from their cylinders directly without purification.

## 3 Electronic Structure Calculations

As reaction is expected to occur over the ground triplet state of the C + $CH_3COCH_3$ system, this PES was investigated at different levels of theory. In the first instance, the geometries and energies of the reagents, intermediates, transition states (TSs) and possible products were calculated using density functional theory (DFT) employing the M06-2X functional[38] coupled with the aug-cc-pVTZ (AVTZ) basis set. This functional has been shown to have good accuracy for the prediction of main-group thermochemistry, kinetics, barrier heights and non-covalent interactions. [39] At the same time, harmonic frequencies were also calculated for these structures at the same level in order to verify the nature of the stationary point (a single imaginary frequency for a TS, no imaginary frequencies for a minimum or product/reagent species). In addition, intrinsic reaction coordinate calculations were performed to confirm that each TS lies on the minimum energy path between 2 assumed minima. For certain critical structures, such as those species that were predicted to lie close in energy to the reagent asymptote at the M06-2X/AVTZ level, we ran more accurate energy calculations based on the geometries obtained at the M06-2X/AVTZ level using the domain based local pair-natural orbital singles and doubles coupled cluster theory with an improved perturbative triples correction algorithm (DLPNO-CCSD(T)). [40] This approach was favoured over standard CCSD(T) calculations due to the large gain in computational time which was further improved by the application of the RIJCOSX approximation. [41] In common with the DFT calculations, the DLPNO-CCSD(T) method was associated with the AVTZ basis set for these calculations. All calculations were performed using ORCA, [42, 43] while the vibrational frequency analysis and structure visualization were performed using Avogadro. [44]

**4 Results**

**4.1 Potential Energy Surface**

A schematic representation of the $^3A''$ surface involved in the C + $CH_3COCH_3$ reaction is shown in Figure 1.

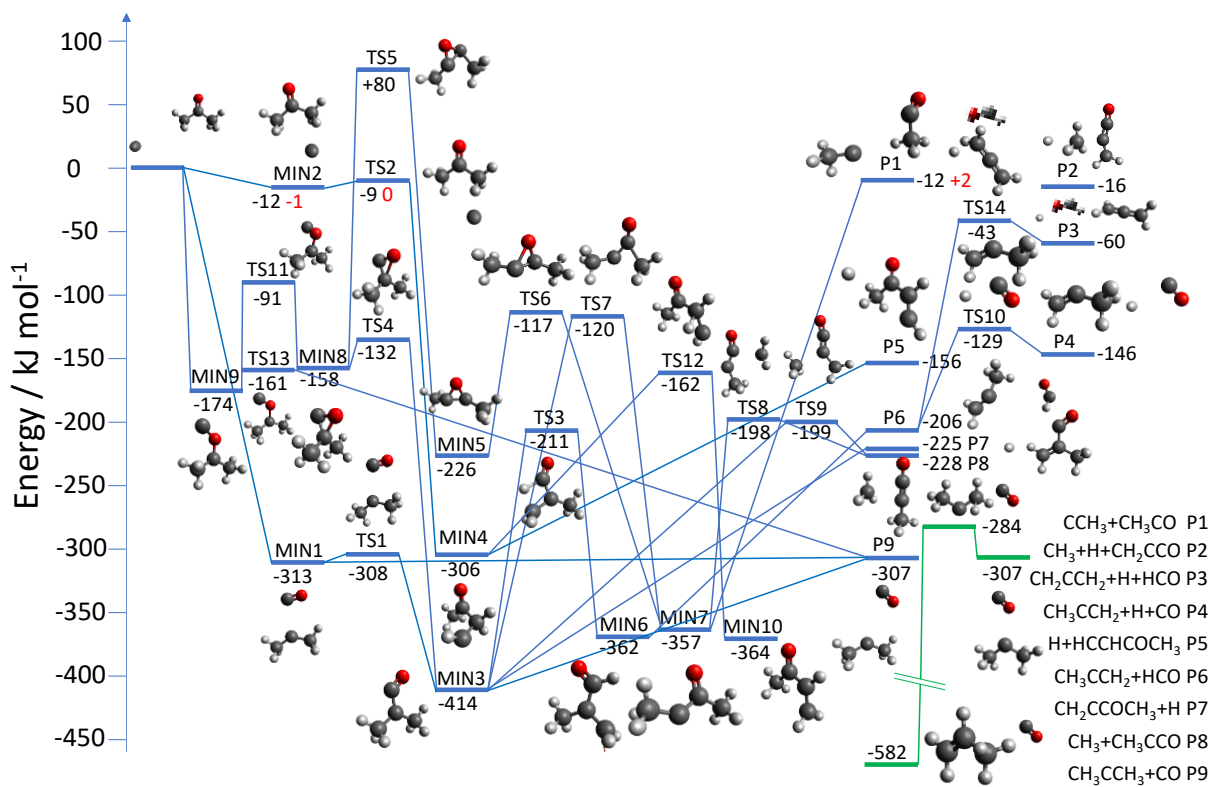

**Figure 1** Schematic diagram of the triplet potential energy surface for the C($^3$P) + CH$_3$COCH$_3$ reaction with energies obtained at the M06-2X/AVTZ level. Certain structures close in energy to the reagent asymptote at this level were calculated with the more accurate DLPNO-CCSD(T)/AVTZ method (energy values in red). Relevant parts of the singlet potential energy surface for the isomerization of $^1$H$_3$CCCH$_3$ to $^1$CH$_3$CHCH$_2$ (propene) are also shown. All energies were corrected for zero-point energy differences.

Energies are displayed at the M06-2X/AVTZ level, with certain critical energies in red displayed at the DLPNO-CCSD(T)/AVTZ level. All energies are corrected for ZPE differences (ZPEs were obtained at the M06-2X/AVTZ level) derived from a vibrational frequency analysis. The green line represents the singlet surface connecting species $^1$CH$_3$CCH$_3$ with $^1$CH$_3$CHCH$_2$ Ten different minima were identified during this work, with nine different product sets connected by fourteen TSs. There are almost certainly several other minima, transition states and product channels that are also energetically accessible, although we believe that we have identified the most important ones to explain the results of our investigation. According to our calculations, ground-state atomic carbon can attack $CH_3COCH_3$ at two different positions in a similar manner to the entrance channels predicted by Joo et al.[45] for the related C + $H_2CO$ reaction.

As a first step, atomic carbon can approach the oxygen atom of the carbonyl group. Here, the C-atom either inserts into the C-O bond of $CH_3COCH_3$ leading to the formation of intermediate species MIN1, 313 kJ/mol below the reagents at the M06-2X/AVTZ level or it adds to the O-atom to form MIN9, -174 kJ/mol below the reagents at the M06-2X/AVTZ level with both approaches being barrierless. MIN1, a weakly bound complex between $CH_3CCH_3$ and CO can either rearrange to MIN3 after overcoming a very low barrier (TS1) of only 5 kJ/mol with respect to MIN1 or dissociate directly to the separated products, $^3CH_3CCH_3$ + $^1CO$ (P9), 307 kJ/mol below the reagents. MIN9 can dissociate to products P9 after passing over a low TS (TS13) only 13 kJ/mol above MIN9, or it can isomerize to the cyclic intermediate MIN8 through a TS 91 kJ/mol below the reagent level (TS11). Once MIN8 is formed, C-O bond breaking can occur (the atoms of the original C=O bond of acetone are those involved here in TS4) to form MIN3. Additional evolution of MIN8 towards another cyclic intermediate MIN5 can only occur over a high barrier, TS5, 238 kJ/mol above MIN8 and 80 kJ/mol above the reagent energy level so this pathway will not occur under the low temperature conditions of the present experiments.

MIN3, potentially formed by both channels MIN1→TS1→MIN3 and MIN9→TS11→MIN8→TS4→MIN3, can isomerize through several energetically accessible pathways to form MIN6 via H-atom transfer over TS3 (213 kJ/mol above MIN3) and MIN7 via $CH_3$ transfer over TS7 (294 kJ/mol above MIN3). MIN3 is also predicted to form bimolecular products through three different pathways. MIN3 can dissociate directly via C-H bond fission to $CH_2CCOCH_3$ +H products (P7), 225 kJ/mol below the reagents or it can dissociate via C-C bond fission to $CH_3$ + $CH_3CCO$ products (P8), 228 kJ/mol below the reagents after passage over an exit TS (TS9) 29 kJ/mol above P8. The C-C bond that breaks here was one of those already present in the $CH_3COCH_3$ molecule. Alternatively, MIN3 can also dissociate directly to products P9 by breaking the newly formed C-C bond.

Once formed, MIN6 can dissociate via C-C bond breaking directly to products $CH_3CCH_2$ + HCO (P6), 206 kJ/mol below the reagents. MIN7 can dissociate to products P8 after fission of the C-C bond between the methyl group and the C-atom of the carbonyl group. This pathway involves an exit TS (TS8) 30 kJ/mol above the P8 asymptote. Alternatively, after fission of the C-C bond on the other side of the carbonyl group, MIN7 can dissociate directly to products $CCH_3$ + $CH_3CO$ (P1), slightly below the reagents at the M06-2X/AVTZ level (-12 kJ/mol). The energy of these products at the DLPNO-CCSD(T)/AVTZ level is +2 kJ/mol (corrected for ZPE), so that this channel may not play any part in the present low temperature experiments. As a third possible pathway, MIN7 can isomerize to MIN5 over TS6, located 240 kJ/mol above MIN7.

For the C + $CH_3COCH_3$ reaction, an alternative pathway exists that is not available in the C + $H_2CO$ reaction.[45] Here, atomic carbon can insert into a C-H bond of one of the methyl groups of $CH_3COCH_3$ to form a weakly bound (-12 kJ/mol at the M06-2X/AVTZ level) van der Waals type complex MIN2. MIN2 can isomerize to MIN4 over TS2, 9kJ/mol below the reagents at the M06-2X/AVTZ level. The calculated energies at the DLPNO-CCSD(T)/AVTZ level are -1 and +0.5 kJ/mol for MIN2 and TS2 respectively, when zero-point-energy corrections (evaluated at the M06-2X/AVTZ level) are taken into account. Given the expected precision of the calculated energies which have been estimated to be within less than 3 kJ/mol of the canonical CCSD(T) results for open-shell systems, [46] it is entirely possible that this represents a non-negligible reaction pathway, particularly at low temperature. After the MIN4 intermediate is formed, this species can isomerize to MIN10 over TS12, 144 kJ/mol above MIN4 or evolve to products H + HCCHCOCH$_3$ (P5) (150 kJ/mol above MIN4) via C-H bond dissociation without a barrier.

**Intersystem crossing**

Interestingly, one of the major products predicted to be formed, $^3$CH$_3$CCH$_3$, could also evolve further in this scenario. This species could undergo intersystem crossing to the singlet state, as $^1$CH$_3$CCH$_3$ (shown in green in Figure 1) is calculated to be at the same energy as the triplet state species at the M06-2X/AVTZ level. Moreover, the geometries of these species are quite similar as the C-C and C-H bond lengths are very close with the only major difference being the C-C-C angles which are 132° and 112° for the triplet and singlet forms respectively, accompanied by different rotational positions of the methyl group hydrogens (see Figure 2).

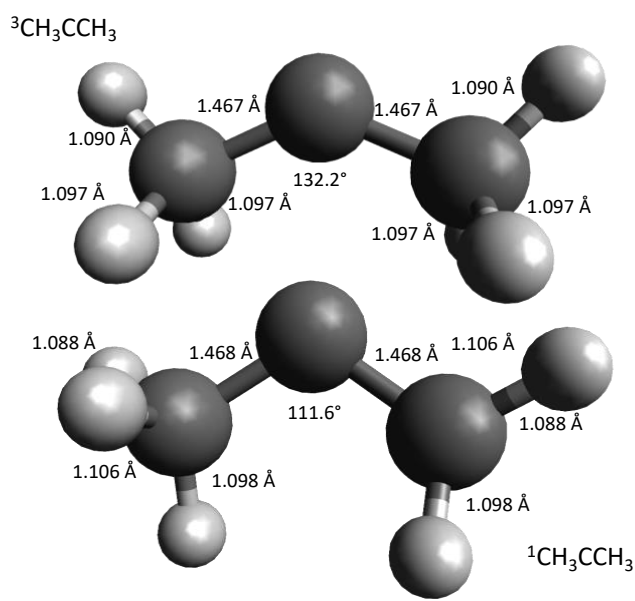

**Figure 2** Calculated geometries of the CH$_3$CCH$_3$ molecule at the M06-2X/AVTZ level in its singlet and triplet state electronic configurations.

In order to check whether intersystem crossing might occur in this case, we ran a minimum energy crossing point (MECP) optimization to determine its energy. The MECP is found to occur only +15 kJ/mol (corrected for ZPE) above the $^3CH_3CCH_3$ minimum at the M06-2X/AVTZ level. Once $^1CH_3CCH_3$ is formed it can isomerize to propene, $C_3H_6$, 275 kJ/mol below $^1CH_3CCH_3$ after overcoming a low TS of +23 kJ/mol relative to $^1CH_3CCH_3$. Considering the energetically favourable pathways towards $C_3H_6$ formation and the possible large branching fraction for the formation of P9 products, if the C + $CH_3COCH_3$ reaction is rapid at low temperature, it could represent a gas-phase source of propene in the dense ISM.

**4.2 Rate constants**

Acetone was present in the flow with a large excess concentration with respect to C($^3P$) during these experiments. Consequently, its concentration was considered to be constant during the individual runs, so it was possible to analyze the kinetics of the C($^3P$) + acetone reaction using a pseudo-first-order treatment. As $CBr_4$ photolysis also produces some C($^1D$) in the flow, it is important to consider whether these atoms might interfere with the rate constant measurements by relaxing back to the ground state or by reacting on the timescale of the experiment. As C($^1D$) is not detected during these experiments while C($^3P$) atoms are detected directly, the expected rapid reaction of C($^1D$) with acetone will not affect the kinetic measurements performed here. Furthermore, as C($^1D$) quenching by Ar is slow,[47] these atoms should not interfere with the rate constant measurements when Ar is used as the carrier gas. When $N_2$ is used as the carrier gas, such as during the experiments conducted at 177 K, C($^1D$) quenching will occur rapidly with a half-life time, $t_{1/2}$, < 1 µS for an $N_2$ flow density of 9.4 × $10^{16}$ cm$^{-3}$. Consequently, the presence of C($^1D$) atoms in the flow are not expected to affect any of the present kinetic measurements. Considering the experimental conditions employed, the C($^3$P) fluorescence signal decays exponentially to zero as a function of time, as shown in Figure 3.

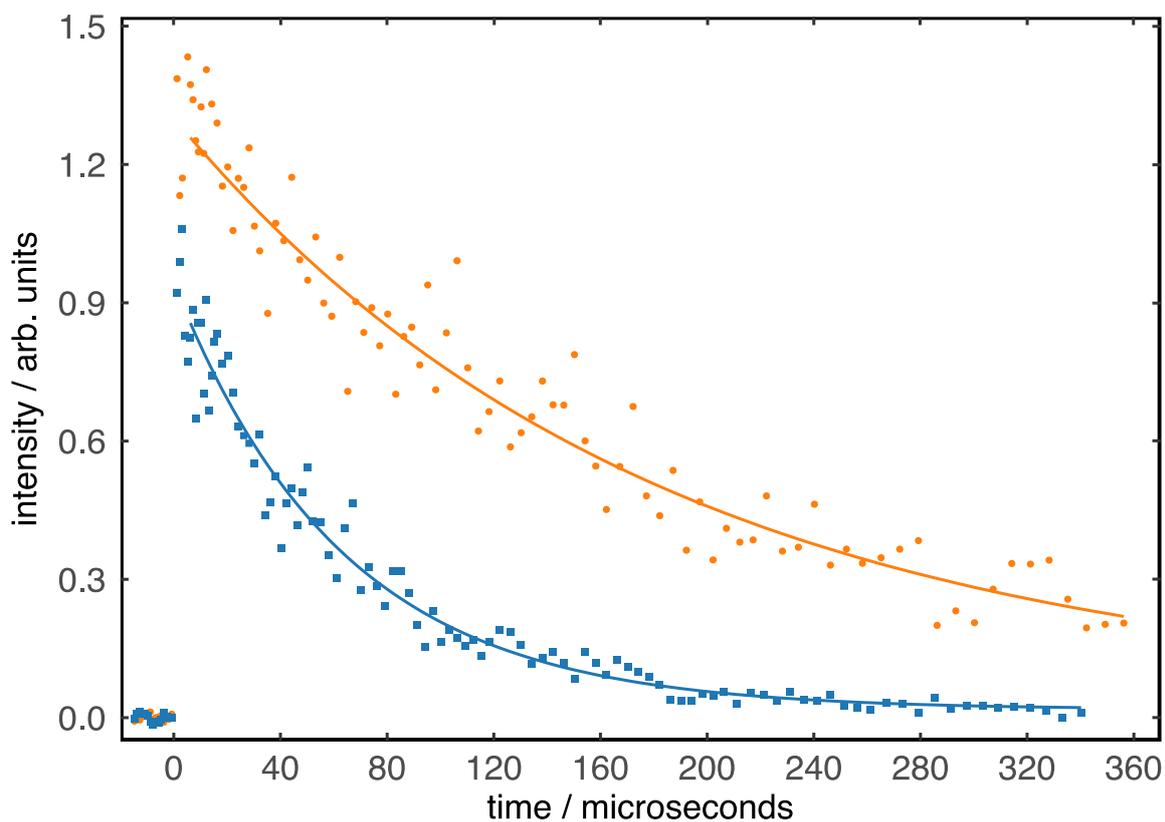

**Figure 3** C($^3$P) fluorescence intensity as a function of delay time recorded at 50 K. (Orange circles) without CH$_3$COCH$_3$; (blue squares) [CH$_3$COCH$_3$] = 4.6 × 10$^{13}$ cm$^{-3}$. Solid lines represent the non-linear least squares exponential fits to the individual datasets.

These decay curves were well-described by an expression of the form

$$I(t) = I_0 \exp(-k_{1st}t) \qquad (1)$$

where $I_0$ is the C($^3$P) fluorescence signal intensity at time zero, $I(t)$ is the C($^3$P) fluorescence signal intensity at time $t$, and $k_{1st}$ is the pseudo-first-order rate constant for atomic carbon loss with units s$^{-1}$. Figure 1 shows that even in the absence of acetone (orange datapoints), the C($^3$P) signal decays exponentially. This occurs due to the physical loss of C-atoms through diffusion out of the zone illuminated by the probe laser (diffusional losses, $k_{diff}$) and by chemical losses such as through the reaction of C-atoms with the precursor molecule CBr$_4$ ($k_{C+CBr_4}[CBr_4]$) while the loss rate increases considerably when acetone is present in the flow (blue datapoints) due to the additional contribution of the C + CH$_3$COCH$_3$ reaction ($k_{C+CH_3COCH_3}[CH_3COCH_3]$). As $k_{diff}$ and $k_{C+CBr_4}[CBr_4]$ are constant for any single series of experiments measuring pseudo-first-order decay rates over a range of [CH$_3$COCH$_3$], a plot of $k_{1st}$ as a function of [CH$_3$COCH$_3$] yields the second-order rate constant $k_{C+CH_3COCH_3}$ for that experiment with units cm$^3$ s$^{-1}$. The individual datapoints were weighted by the uncertainty derived during the non-linear least squares fitting procedure used to obtain the first order rate constants, according to the expression w = 1/σ$^2$ (w is the weight, σ is the uncertainty) so that the data points with the largest associated uncertainty carried the least weight. Several of these second-order plots recorded at different temperatures over the 50-296 K range can be seen in Figure 4.

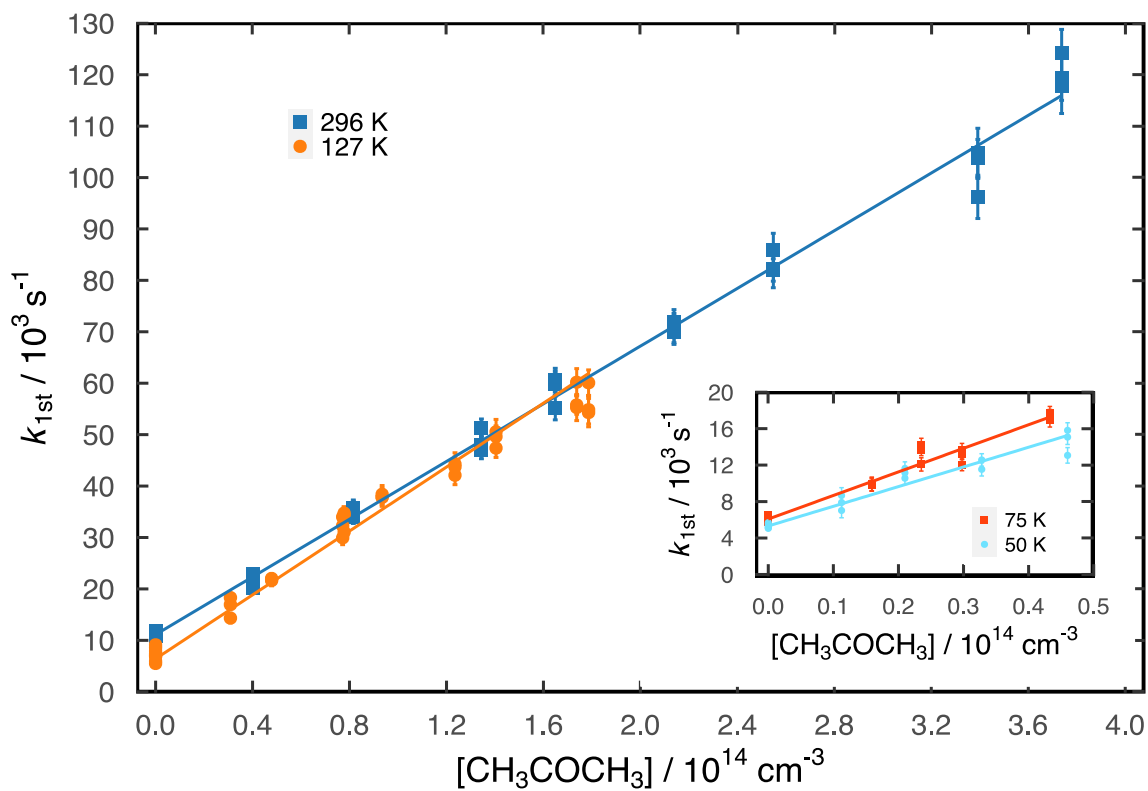

**Figure 4** Pseudo-first-order rate constants for atomic carbon loss as a function of the acetone concentration. (Dark blue squares) data recorded at 296 K; (orange circles) data recorded at 127 K. **Inset** (red squares) data recorded at 75 K; (light blue circles) data recorded at 50 K. Solid lines represent weighted linear least-squares fits to the data. Error bars represent the uncertainties on individual pseudo-first-order rate constants derived from non-linear fits to the atomic carbon fluorescence profiles such as those shown in Figure 3.

At least five different concentration values were used to derive the second-order rate constants at any single temperature, based on a minimum of 17 datapoints. The intercept values of these plots at the abscissa correspond to the sum of the $k_{\text{diff}} + k_{\text{C+CBr}_4}[\text{CBr}_4]$ contributions to the $k_{\text{1st}}$ values. At the lowest temperatures (50 and 75 K), the derived $k_{\text{1st}}$ values only varied linearly as a function of the acetone concentration over a limited range.

Beyond these limiting values, $k_{1st}$ was seen to increase more slowly as the CH₃COCH₃ concentration increased. This observation was considered to be a sign of acetone cluster formation within the cold supersonic flow, effectively preventing a fraction of the acetone monomers from reacting with atomic carbon and lowering the expected pseudo-first-order rate constant. Additionally, the reaction between C($^3$P) and acetone dimers, trimers, etc… may not occur at the same rate as the equivalent monomer reaction and/or lead to alternative reaction pathways. Consequently, only those experiments employing [CH₃COCH₃] < 5 × 10¹³ cm⁻³ were used in the final analysis for measurements recorded at 50 and 75 K. The derived second-order rate constants are summarized in Table 1 alongside other relevant information such as the number of measurements and the concentration ranges used.

**Table 1** Measured second-order rate constants for the C($^3$P) + CH₃COCH₃ reaction

| T / K | $N^b$ | [CH₃COCH₃] / 10¹³ cm⁻³ | Flow density] / 10¹⁷ cm⁻³ | $k_{C(^3P)+CH_3COCH_3}$ / 10⁻¹⁰ cm³ s⁻¹ | Carrier gas |
|---|---|---|---|---|---|
| 296 | 30 | 0 - 37.4 | 1.65 | (2.81 ± 0.28)$^c$ | Ar |
| 177 ± 2$^a$ | 30 | 0 - 12.9 | 0.94 | (2.72 ± 0.28) | N₂ |
| 127 ± 2 | 36 | 0 - 17.9 | 1.26 | (3.11 ± 0.32) | Ar |
| 75 ± 2 | 18 | 0 - 4.3 | 1.47 | (2.59 ± 0.28) | Ar |
| 50 ± 1 | 17 | 0 - 4.6 | 2.59 | (2.16 ± 0.24) | Ar |

$^a$Uncertainties on the calculated temperatures represent the statistical (1σ) errors obtained from Pitot tube measurements of the impact pressure. $^b$Number of individual measurements. $^c$Uncertainties on the measured rate constants represent the combined statistical (1σ) and estimated systematic errors (10%).

These values as well as those determined in earlier work are displayed as a function of temperature in Figure 5.

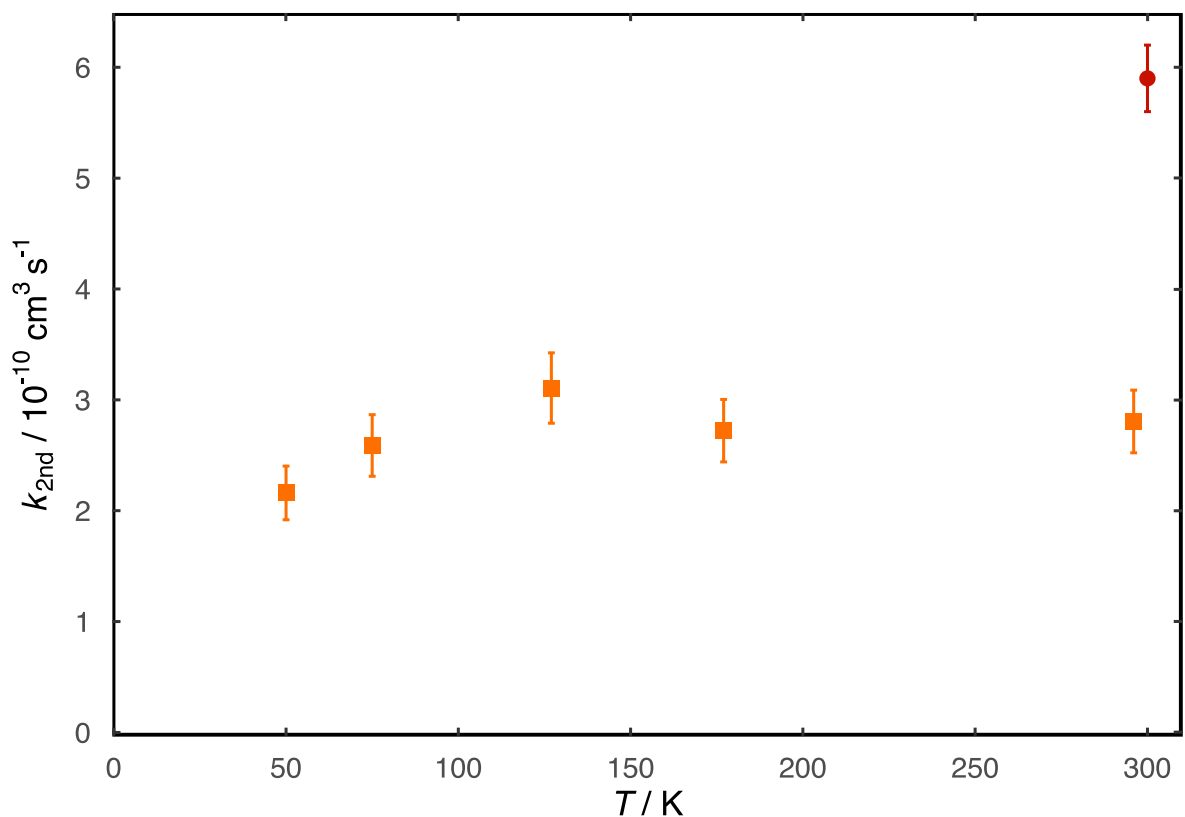

**Figure 5** Rate constants for the C + $CH_3COCH_3$ reaction as a function of temperature. (Red circles) Husain and Ioannou;[48] (orange squares) this work. Error bars on the present values represent the combined statistical and systematic uncertainties. Statistical uncertainties were derived from weighted linear-least squares fits to the pseudo-first-order rate constants plotted as a function of the acetone concentration as shown in Figure 4. Systematic uncertainties were estimated to be 10 % of the nominal value of the second-order rate constant.

The measured room temperature rate constant is large, with a value $k_{\text{C}+\text{CH}_3\text{COCH}_3}(296\text{ K})$ = (2.81 ± 0.28) × 10⁻¹⁰ cm³ s⁻¹ that is consistent with the presence of at least one barrierless pathway for the C + CH₃COCH₃ reaction. The only other previous kinetic study of the C + CH₃COCH₃ reaction was performed at room temperature by Husain and Ioannou,[48] who derived a rate constant $k_{\text{C}+\text{CH}_3\text{COCH}_3}(300\text{ K})$ = (5.9 ± 0.3) × 10⁻¹⁰ cm³ s⁻¹; a value more than twice as large as the value derived here at room temperature. Husain and Ioannou[48] used a conventional flash photolysis apparatus to study this process. In their work, C(³P) atoms were formed by the VUV photolysis (λ > 160 nm) of carbon suboxide, C₃O₂, using a coaxial lamp. Previous work has demonstrated that the major photolysis product in this wavelength range is ground state C(³P) atoms rather than excited state C(¹D) atoms (97 % and 3 % respectively) so that relaxation of C(¹D) atoms back to the ground state was expected to be essentially negligible during this work. C(³P) atoms were followed by absorption, using the emission from an atomic resonance lamp at 166 nm, optically isolated using a VUV monochromator and a solar blind PMT. The large difference in rate constant values between the present work and this previous study is difficult to understand. Nevertheless, it is clear that the use of an absorption-based detection method (as opposed to the fluorescence-based method used here) requires the use of much larger C(³P) concentrations for equivalent signal-to-noise ratios, potentially leading to higher levels of secondary reactions. Moreover, as the C-atom precursor C₃O₂ has a very large absorption cross-section around 166 nm (~1 × 10⁻¹⁶ cm²), any concentration changes of this species during the course of a single experiment (such as through depletion by the repetitive photolysis technique employed by Husain and Ioannou[48]) would alter the transmitted signal levels leading to an incorrect decay profile. As the C-atom precursor molecule used in our study (CBr₄) has an absorption cross-section at 266 nm that is a hundred times smaller than the $C_3O_2$ cross-section at 166 nm, the equivalent issue should not arise in the present work.

The measured rate constants show only small variations as a function of temperature, with values of (2.1-3.1) × $10^{-10}$ cm$^3$ s$^{-1}$ over the 50-296 K range. The absence of a strong temperature dependence of the rate constant would seem to indicate that the major pathways are not strongly influenced by the presence of a pre-reactive complex in the entrance channel. Indeed, earlier studies of the barrierless reactions of C($^3$P) with other reagents involving complex formation and a submerged barrier towards adduct formation[5, 14, 31] present much stronger temperature dependences than those C-atom reactions where these features are absent.[15] Despite this, the measured reaction rates at 75 and 50 K are slightly lower than those obtained at higher temperature, although the differences are close to being covered by the combined measurement uncertainties. In this respect, for low temperature interstellar modelling purposes, we recommend the use of a temperature independent value for the rate constant, $k_{C+CH_3COCH_3} = 2.2 \times 10^{-10}$ cm$^3$ s$^{-1}$. Mechanistically, the observed rate constant decrease could be indicative of the opening of the reactive channel via TS2 as the temperature falls (see Figure 1), with part of the reactive flux being trapped in MIN2 rather than passing by the other barrierless channels via MIN1 and MIN9. Further evidence for the opening of this channel will be presented in section 4.3.

## 4.3 Temperature dependent H-atom yields

The measurement of H-atom VUV LIF intensities at 75, 177 and 296 K allowed us to gain some insight into the H-atom forming product channels of the C + CH$_3$COCH$_3$ reaction. In order to put these measurements on an absolute scale, the derived intensity was compared to the one recorded for a reference process. In this case, the C + C$_2$H$_4$ → C$_3$H$_3$ + H reaction, with a measured H-atom yield of 0.92 ± 0.04 at 300 K,[49] was used to calibrate the H-atom yield of the C + $CH_3COCH_3$ reaction. We assume that this yield does not change over the present temperature range due to the absence of a prereactive complex for this process coupled with the absence of any significant (submerged) barriers over the triplet PES leading to $C_3H_3$ + H products.[50] As $C(^1D)$ atoms were also present in the supersonic flow during these experiments, it was necessary to check that the production of any additional H-atoms by the reactions of $C(^1D)$ with both $CH_3COCH_3$ and $C_2H_4$ did not interfere with our measurements of the equivalent ground state processes. These checks were performed by replacing all (or part) of the carrier gas Ar by $N_2$ in some experiments, as $N_2$ is known to quench $C(^1D)$ atoms much more rapidly than Ar, with a quenching rate constant that increases as the temperature falls.[32] In this respect, experiments were carried out at 296 K using both Ar and $N_2$ as the carrier gases, while a large concentration of $N_2$ ($1.5 \times 10^{16}$ $cm^{-3}$) was added to the flow during measurements performed at 75 K. The carrier gas used to produce supersonic flows at 177 K is $N_2$, so no additional checks were required for these measurements.

Unfortunately, as can be seen from Figure 6, recorded at 75 K, the present experiments were hampered by the production of atomic hydrogen from the photolysis of acetone at 266 nm which added to the H-atom signal produced by the C + $CH_3COCH_3$ reaction.

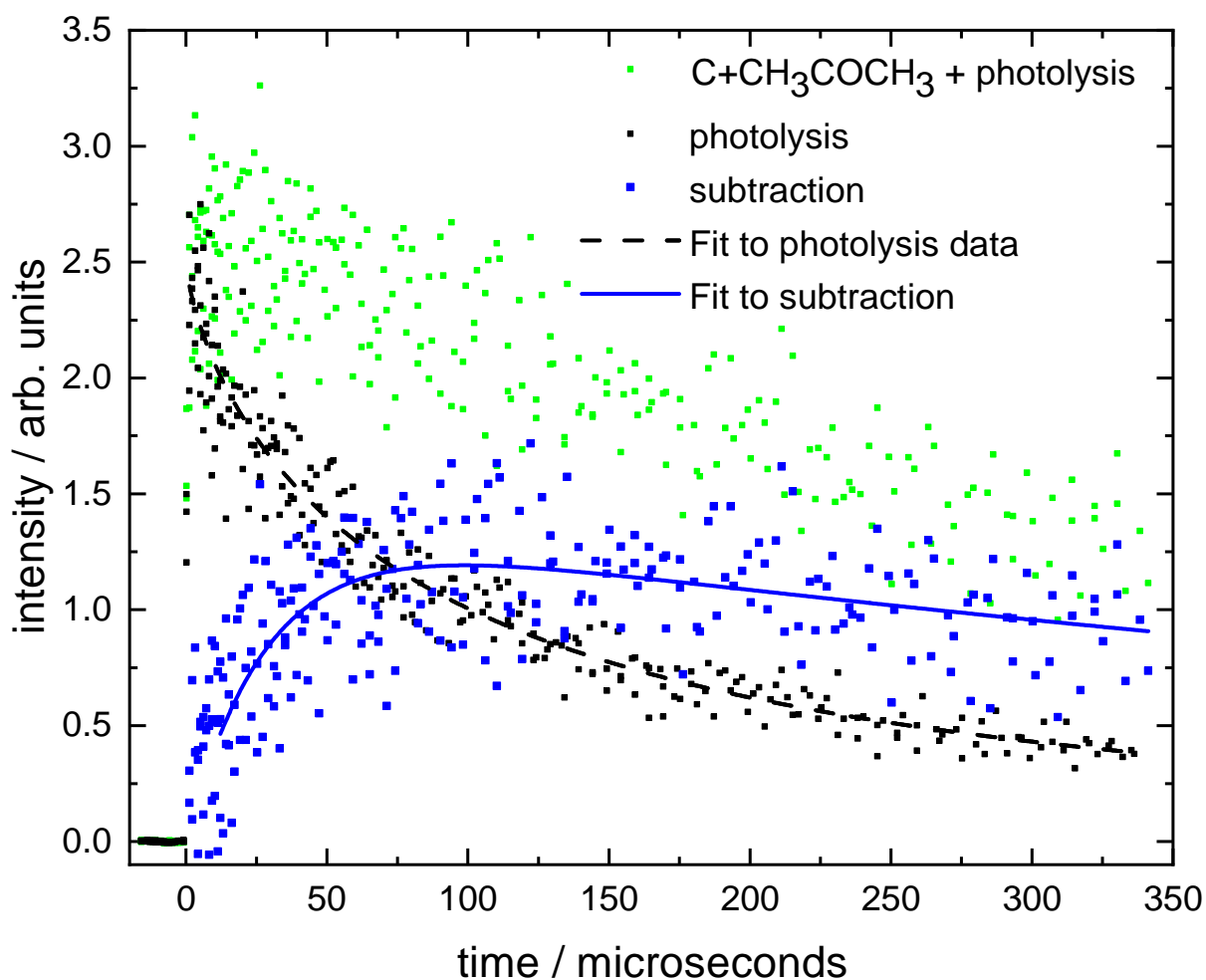

**Figure 6** Recorded H-atom signals as a function of time at 75 K. (Solid green squares) H-atom signal produced by $CH_3COCH_3$ photolysis and the C + $CH_3COCH_3$ reaction. (Solid black squares) H-atom signal produced by $CH_3COCH_3$ photolysis alone (no $CBr_4$ was present in the flow). (Dashed black line) fit to the $CH_3COCH_3$ photolysis data. (Solid blue squares) subtraction of the photolysis contribution from the green datapoints, representing the reactive H-atom signal. (Solid blue line) biexponential fit to the reactive H-atom signal.

In order to extract the contribution from the C + $CH_3COCH_3$ reaction alone, two different experiments were performed. Firstly, the H-atom temporal signal was recorded with the photolysis laser on, with both atomic carbon and $CH_3COCH_3$ present in the supersonic flow to yield the sum of the two contributions (reaction + photolysis - green points in Figure 6). Then, a second experiment was performed where the vessel containing $CBr_4$ was isolated from the supersonic flow. Under these conditions, no C-atoms are present in the flow so that only the H-atom signal from acetone photolysis is recorded (black points in Figure 6). This procedure is repeated at least 3 or 4 times and the resulting curves of the same type are coadded and analyzed. This method was preferred over the one used in previous work, where pairs of curves (target reaction followed by reference reaction or vice-versa) were recorded and analyzed separately, due to the larger uncertainty of the present work brought about by the requirement to subtract the H-atom signal generated by photolysis. Indeed, Figure 6 above recorded at 75 K shows the results of three coadded temporal curves of each type (photolysis+reaction and photolysis alone).

An arbitrary function is chosen to produce a good fit to the photolysis data, which is then subtracted from the trace containing the photolysis+reactive contribution to leave the reactive contribution alone (blue points in Figure 6). It can be seen that these data have the biexponental form expected for a production process in Laval nozzle flows (H-atom formation followed by its diffusional loss). Consequently, these data were well described by the following expression

$$I_H = A\{\exp(-k_{L(H)}t) - \exp(-k_{1st}t)\} \qquad (2)$$

to yield the blue line shown in Figure 6. Here, $I_H$ is the H-atom signal intensity and $k_{L(H)}$ represents the secondary H-atom loss through processes such as diffusion. These data are then compared to the H-atom signal produced by the $C + C_2H_4$ reference reaction as shown in Figure 7 for two different temperatures (upper panel - 296 K, lower panel - 75 K).

S25

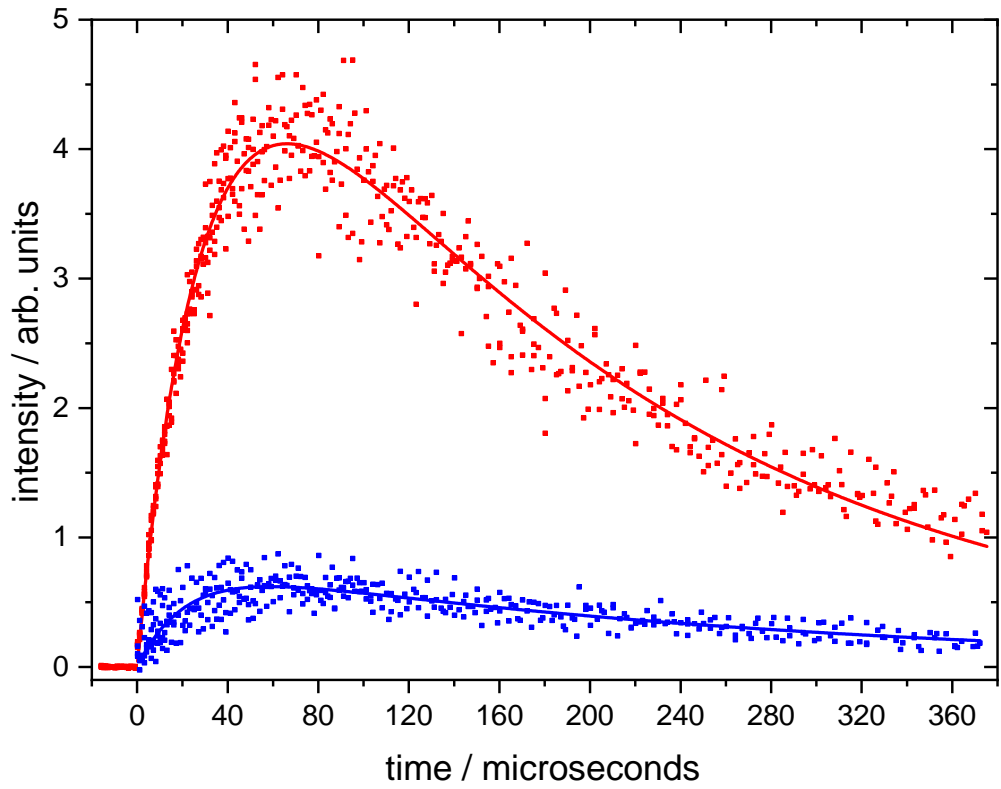

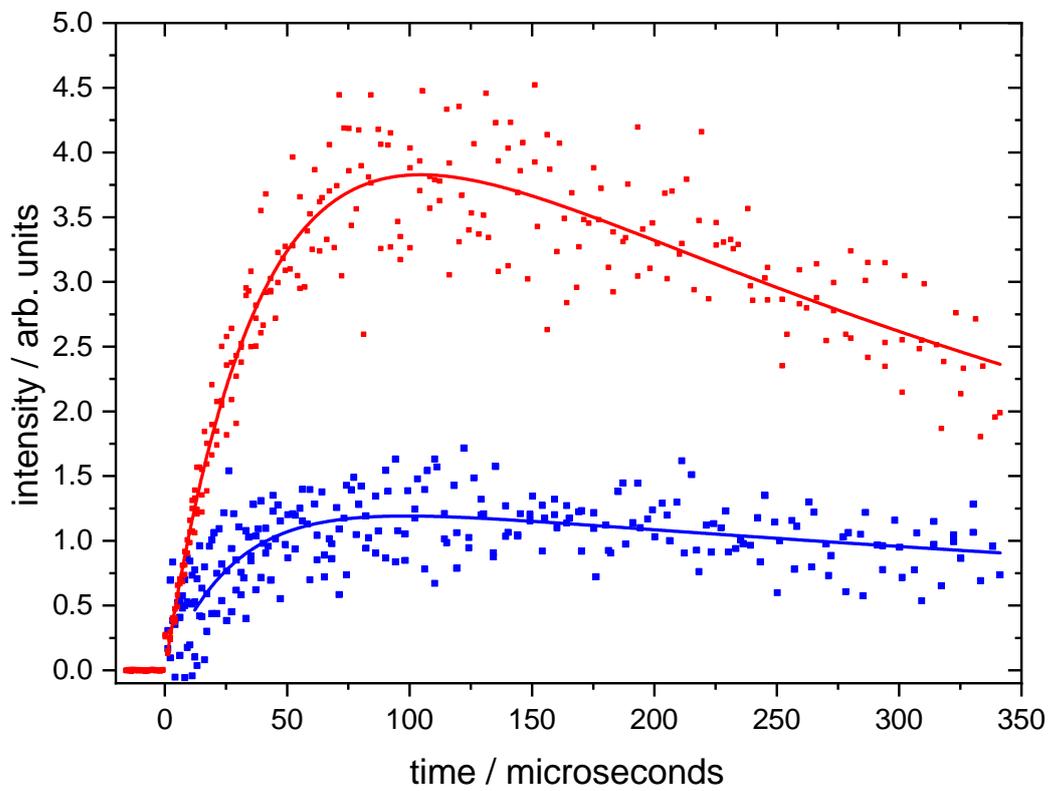

**Figure 7** (Upper panel) H-atom formation curves recorded at 296 K. (red solid squares) H-atom formation by the C + $C_2H_4$ reaction ([$C_2H_4$] = 4.3 × $10^{13}$ $cm^{-3}$). (Blue solid squares) H-atom formation by the C + $CH_3COCH_3$ reaction ([$CH_3COCH_3$] = 2.4 × $10^{13}$ $cm^{-3}$) resulting from the subtraction procedure described in the text and displayed in Figure 6. (Lower panel) H-atom formation curves recorded at 75 K. (red solid squares) H-atom formation by the C + $C_2H_4$ reaction ([$C_2H_4$] = 1.5 × $10^{13}$ $cm^{-3}$). (Blue solid squares) H-atom formation by the C + $CH_3COCH_3$ reaction ([$CH_3COCH_3$] = 1.7 × $10^{13}$ $cm^{-3}$) resulting from the subtraction procedure described in the text and displayed in Figure 6. Solid red and blue lines represent biexponential fits to the subtracted data.

To be consistent with the data for the C + $CH_3COCH_3$ target reaction, all H-atom temporal curves for the C + $C_2H_4$ reference reaction were also co-added before analysis. The peak intensities of the H-atom formation curves for the C + $C_2H_4$ reference reaction shown in Figure 7 were then divided by 0.92 to correct for the fact that the measured H-atom yields for this process are slightly smaller than 1. [49] Absolute H-atom yields for the $C(^3P)$ + $CH_3COCH_3$ reaction were then obtained by dividing the peak values from the fitted curves for the $C(^3P)$ + $CH_3COCH_3$ reaction by the corrected peak values for the $C(^3P)$ + $C_2H_4$ reaction. No further corrections were required in the present experiments as the low $CH_3COCH_3$ and $C_2H_4$ concentrations used here were estimated to result in less than 1 % absorption losses at 121.567 nm. The derived absolute H-atom yields for the $C(^3P)$ + $CH_3COCH_3$ reaction are listed in Table 2.

**Table 2** H atom yields for the $C(^3P)$ + $CH_3COCH_3$ reaction as a function of temperature

| T / K | Number of experiments | H atom yield |
|-------|----------------------|--------------|
| 296   | 12                   | 0.13 ± 0.02  |
| 177   | 7                    | 0.25 ± 0.04  |
| 75    | 8                    | 0.29 ± 0.03  |

[a]The error bars reflect the statistical uncertainties at the level of a single standard deviation including the uncertainties of the H branching ratio of the C + $C_2H_4$ reaction used as a reference with a systematic error of 10 % added to account for the higher uncertainty brought about by the subtraction procedure described above.

It can be clearly seen from Figure 7 and Table 2 that the H-atom yield of the C + $CH_3COCH_3$ reaction becomes more important, relative to the yield of the C + $C_2H_4$ reaction, at low temperature although H-atom formation pathways are minor channels at all temperatures. Considering the exothermicity of MIN1 followed by the direct dissociative pathway to products P9, it is likely that this represents the major exit channel of the C + $CH_3COCH_3$ reaction at room temperature and below. Nevertheless, we cannot exclude the transitory formation of MIN3, given the low value of TS1 and the high exothermicity of MIN3. We would then expect that MIN3 dissociates to the most thermodynamically stable products P9 through C-C bond fission although several other channels such as the ones towards P7 (H + $CH_2COCH_3$) through C-H bond fission and P8 ($CH_3$ + $CH_3CCO$) through C-C fission, and the isomerization pathway towards MIN6 and eventually P6 (HCO + $CH_3CCH_2$) are also energetically accessible so these cannot be excluded. The pathways MIN9→TS11→MIN8→TS4→MIN3 and MIN9→TS13→P9 might also play a role in the reaction. TS13 is 161 kJ/mol below the reagent level, so this latter pathway should not present any temperature dependence. Although TS11 is only 91 kJ/mol below the reagents, it also seems unlikely that this would be high enough to lead to significant temperature dependent effects in the former pathway. The main predicted pathways of the C + CH$_3$COCH$_3$ reaction (that is via MIN1→P9 and MIN1→MIN3→P9) do not present any substantial barriers (or submerged barriers) to product formation. As the temperature falls it is expected that the lifetime of MIN3 will increase which might favour the formation of products other than CH$_3$CCH$_3$ + CO. Indeed, as no TS was located along the dissociation pathway MIN3→P7 (CH$_2$CCOCH$_3$+H), this pathway may represent a minor exit channel producing H-atoms. Similarly, the pathways MIN3 → MIN6 → P6(HCO+CH$_3$CCH$_2$) → P4(H+CO+CH$_3$CCH$_2$) and MIN3 → MIN6 → P6(HCO+CH$_3$CCH$_2$) → P3(H+HCO+H$_2$CCCH$_2$) could also be sources of H-atoms as both HCO and CH$_3$CCH$_2$ dissociation to H + CO and H + H$_2$CCCH$_2$ respectively are energetically accessible exit channels. To test for the possibility of HCO dissociation to H + CO, experiments were conducted at room temperature with two different densities of N$_2$ carrier gas (1.6 and 3.2 × 10$^{17}$ cm$^{-3}$). According to the study by Langford and Moore, [51] the vibrational relaxation process HCO(0,1,0) + N$_2$ → HCO(0,0,0) + N$_2$ is characterized by a rate constant of 2.5 × 10$^{-13}$ cm$^3$ s$^{-1}$ leading to pseudo-first-order decay rates of 4 and 8 × 10$^4$ s$^{-1}$. Earlier studies of the unimolecular dissociation of HCO over a range of pressures and temperatures[52] indicate that this process is likely to be an order of magnitude slower (a few 10$^3$ s$^{-1}$) so if H-atoms were produced predominantly by this mechanism we would expect to see a notable difference in the H-atom yields as a function of pressure. The H-atom yields derived by these two experiments were almost identical (0.13 at lower pressure compared with 0.14 at higher pressure), indicating that H-atom formation by HCO dissociation is likely to be a minor contribution overall.

Alternatively, the pathway MIN2→MIN4→P5 might become more important as the temperature falls, even though the energies calculated at the DLPNO-CCSD(T)/AVTZ energies are close to the reagent level. Here, the formation of the weakly bound van der Waals type complex MIN2 (-1 kJ/mol at the DLPNO-CCSD(T)/AVTZ level corrected for ZPE) should become more favorable at low temperature, diverting a fraction of the reactive flux away from the major pathways described above. As TS2 is also low (0 kJ/mol at the DLPNO-CCSD(T)/AVTZ level corrected for ZPE), the production of P5 (H + HCCHCOCH$_3$) via MIN4 could explain the observed increase in H-atom production as the temperature falls. In order to validate this hypothesis, additional electronic structure calculations should be performed at a higher level of theory (such as by using double-hybrid DFT or wavefunction based methods for the geometry optimization steps and/or larger basis sets) coupled with statistical RRKM type calculations to provide further interpretation of the experimental data, with H-atom formation potentially arising from several different product channels (MIN3→MIN6→P6→P4, MIN3→MIN6→P6→P3, MIN3→P7, MIN2→MIN4→P5). Nevertheless, if the barrier height TS2 for the channel MIN2→MIN4→P5 is submerged with respect to the reagent level, this pathway represents a plausible explanation for the observed increased H-atom yield at lower temperature. As an alternative possibility, as mentioned above, the longer expected lifetime of MIN3 as the temperature falls could also favour the production of H-atoms through the H + CH$_2$CCOCH$_3$ exit channel (P7).

**5 Astrochemical Model and Comparison with Observations**

Considering the absence of the C + CH$_3$COCH$_3$ reaction in astrochemical databases such as KIDA, [22] it is interesting to test the effect of this process on the abundances of CH$_3$COCH$_3$ and other related species predicted by astrochemical models. Here, we used the gas-grain model Nautilus[53, 54] in its three-phase form[55, 56] to simulate the abundances of atoms and molecules in neutral and ionic form as a function of time, employing kida.uva.2014[22] as the basic reaction network updated recently for a better description of COMs on grains and in the gas-phase.[18, 57, 58] 800 individual species are included in the network that are involved in 9000 separate reactions. Elements are either initially in their atomic or ionic forms in this model (elements with an ionization potential < 13.6 eV are considered to be fully ionized) and the C/O elemental ratio is equal to 0.71 in this work. The initial simulation parameters are listed in Table 3.

**Table 3** Elemental abundances and other model parameters

| Element | Abundance[a] | nH + 2nH$_2$ / cm$^{-3}$ | T/ K | Cosmic ray ionization rate / s$^{-1}$ | Visual extinction |
|---|---|---|---|---|---|
| H$_2$ | 0.5 | $2.5 \times 10^4$ | 10 | $1.3 \times 10^{-17}$ | 10 |
| He | 0.09 | | | | |
| C$^+$ | $1.7 \times 10^{-4}$ | | | | |
| N | $6.2 \times 10^{-5}$ | | | | |
| O | $2.4 \times 10^{-4}$ | | | | |
| S$^+$ | $1.5 \times 10^{-5}$ | | | | |
| Fe$^+$ | $3.0 \times 10^{-9}$ | | | | |
| Cl$^+$ | $1.0 \times 10^{-9}$ | | | | |
| F | $6.7 \times 10^{-9}$ | | | | |

[a]Relative to total hydrogen (nH + 2nH$_2$)

The grain surface and the mantle are both chemically active for these simulations, while accretion and desorption are only allowed between the surface and the gas-phase. The dust-to-gas ratio (in terms of mass) is 0.01. A sticking probability of 1 is assumed for all neutral species while desorption can occur by thermal and non-thermal processes (cosmic rays, chemical desorption) including sputtering of ices by cosmic-ray collisions.[59] The surface reactions formalism and a more detailed description of the simulations can be found in Ruaud et al.[55]

A large fraction of the CH$_3$COCH$_3$ present in our updated model[18] originates from the reactions of the C$_3$ species. Here, the C$_3$ + H$_3^+$ reaction, followed by a sequence of hydrogen transfers with H$_2$ lead eventually to the formation of C$_3$H$_5^+$ and C$_3$H$_7^+$. Then, the OH + C$_3$H$_7^+$ and $H_2O + C_3H_5^+$ reactions produce $CH_3COHCH_3^+$ (with a secondary contribution from the $C_2H_4$ + $H_2COH^+$ reaction) which undergoes electronic Dissociative Recombination (DR) to yield $CH_3COCH_3$.

The neutral O + 2-$C_3H_7$ ($CH_3$-CH-$CH_3$) reaction,[60] where 2-$C_3H_7$ itself is produced on grains, is also a minor but non-negligeable route for producing $CH_3COCH_3$. The main losses of gas-phase $CH_3COCH_3$ are through reaction with $H_3^+$ and $HCO^+$, through reaction with atomic C (this work) and the OH radical,[25, 61] and by depletion onto grains.

As can be seen from Figure 8, introduction of the C + $CH_3COCH_3$ reaction into the network with an estimated rate constant, $k_{C+CH_3COCH_3}(10\ K)$ = 2.2 × 10$^{-10}$ cm$^3$ s$^{-1}$, leads to a maximum decrease of the $CH_3COCH_3$ abundance of up to two orders of magnitude between 10$^3$ and 10$^5$ years. For the present test, we chose to limit product formation to the $C_3H_6$ + CO channel as the possible $C_4H_5O$ product isomers in particular are not currently included in the model.

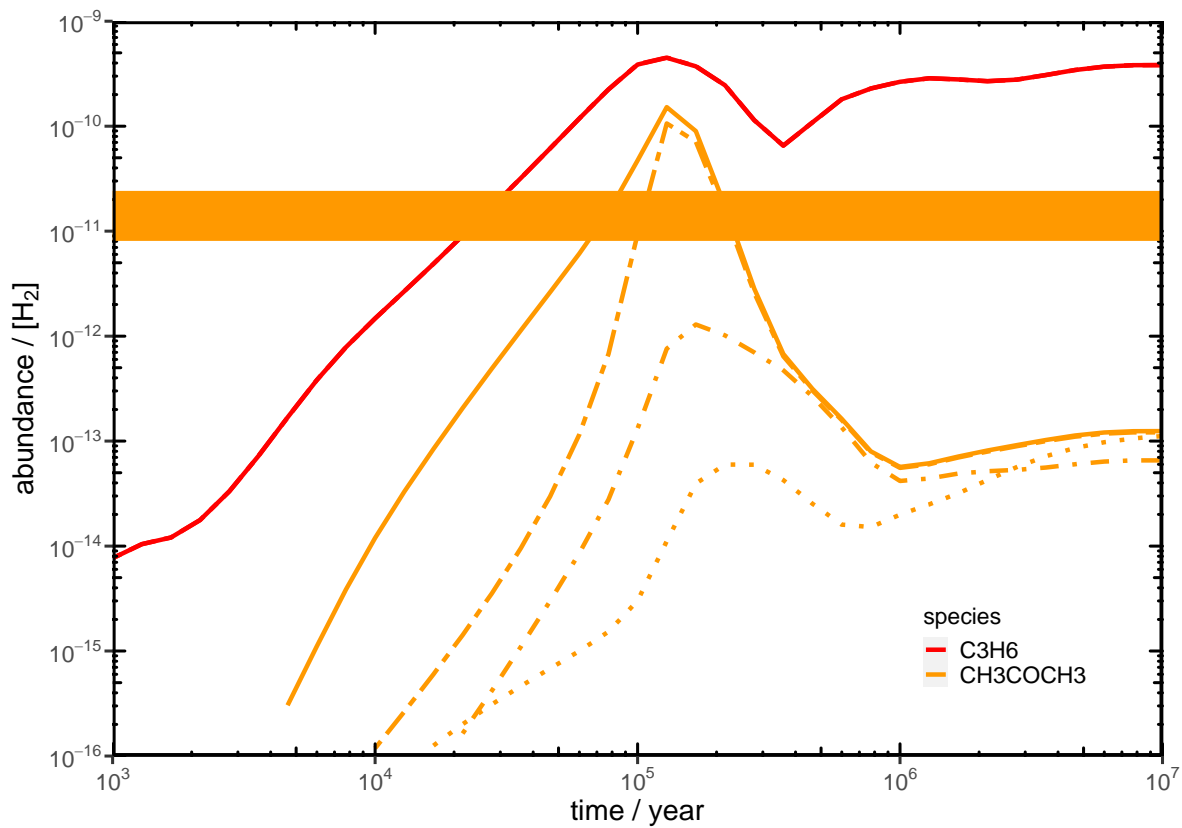

**Figure 8** Gas-grain astrochemical model results for the formation of $CH_3COCH_3$ (orange lines) and $C_3H_6$ (red lines) in dark clouds as a function of cloud age. (Solid line) standard network results. (Short dashed - long dashed line) the same network but with the gas-phase C + $CH_3COCH_3$ reaction included. (Dashed-dotted line) new network without the l-$C_3H_3^+$ + $H_2$ and $C_3H_5^+$ + $H_2$ reactions. (Dotted line) new network with the rate constant of the gas-phase O + $C_3$ reaction set to $1 \times 10^{-11}$ cm$^3$ s$^{-1}$. The $C_3H_6$ abundances given by the standard network (Red solid line) and the network including the C + $CH_3COCH_3$ reaction (Red short dashed - long dashed line) are indistinguishable on the present scale. The horizontal orange rectangle represents the observed $CH_3COCH_3$ abundance in TMC-1[18] with an arbitrary error associated ($\pm\sqrt{3}$).

At ages considered to be characteristic of typical dense clouds (a few $10^5$ years), atomic carbon is removed from the gas-phase through gas-phase reactions forming CO and by accretion onto grains, thereby limiting the effect of the C + $CH_3COCH_3$ reaction at longer times (>$10^5$ years). Despite the effect of the C + $CH_3COCH_3$ reaction on $CH_3COCH_3$ abundances at early times, these simulations indicate that the C + $CH_3COCH_3$ reaction induces only small changes in the gas-phase abundance of $CH_3COCH_3$ at typical dense interstellar cloud ages, with a calculated $CH_3COCH_3$ abundance in good agreement with the observations for TMC-1[18] at ages considered to be characteristic of typical dense clouds (a few $10^5$ years). As $C_3H_6$ is the expected major product of the C + $CH_3COCH_3$ reaction, the predicted $C_3H_6$ abundances for our standard network and for the network including the C + $CH_3COCH_3$ reaction are displayed in red in Figure 8. These two curves essentially overlap, indicating that the title reaction is only a very minor source of gas-phase propene in this model. The major source of $C_3H_6$ in our simulations is the $C_3H_7^+$ + $e^-$ DR reaction.

    As discussed above, according to our updated network, the chemistry of acetone in dense clouds is strongly linked to the chemistry of $C_3$, which provides the $C_3$ backbone for this molecule. The agreement between the observations and the simulations for $CH_3COCH_3$ is therefore highly dependent on the efficiency of the O + $C_3$ reaction, for which the calculations by Woon and Herbst[62] predict a barrier in the entrance channel, leading to a negligible reaction rate at 10 K. This agreement is also highly dependent on the rates of the reactions of $C_3H_3^+$ and $C_3H_5^+$ with $H_2$ which are very slow at room temperature[63] but which can be non-negligible at low temperature[64] due to tunneling. The effects of changing the reactivities of $C_3$ with atomic oxygen, and of $C_3H_3^+$ and $C_3H_5^+$ with $H_2$ are shown in Figure 8. If $C_3$ is allowed to react with atomic oxygen at 10 K, $CH_3COCH_3$ is largely underestimated in the model. However, if O and $C_3$ are unreactive at low temperature, it allows the observed $CH_3COCH_3$

abundance in dense molecular clouds to be reproduced but induces a high abundance of $C_3$ in the gas-phase which seems to be incompatible with the $^{13}C$ fractionation of c-$C_3H_2$ [65] and with the observed abundances of $CH_3CCH$ and $C_3H_6$ in protostars.[57] If the rates of the reactions of $C_3H_3^+$ and $C_3H_5^+$ with $H_2$ are negligible at 10 K, the model significantly underestimates the $CH_3COCH_3$ abundance as these processes control the abundances of $C_3H_5^+$ and $C_3H_7^+$ which are important species for the production of $CH_3COCH_3$. In parallel, these hydrogenation reactions also allow the model to reproduce the observations of $CH_3CCH$[66] and $C_3H_6$,[67] which models struggle to reproduce otherwise.[68] It should also be noted that in addition to $CH_3CCH$, $C_3H_6$ and $CH_3COCH_3$, the reactions of $C_3H_3^+$ and $C_3H_5^+$ with $H_2$ also control the production of several other species including $CH_3CHO$, produced mainly by the O + $C_2H_5$ reaction. $CH_3CHO$ is largely underestimated in the models if $C_3H_6$ is also underestimated because the main source of $C_2H_5$ radicals in dense molecular cloud networks is the O + $C_3H_6$ reaction.[69, 70] Clearly, considering the number of species affected by the O + $C_3$ reaction and the reactions of $C_3H_3^+$ and $C_3H_5^+$ with $H_2$, including $CH_3COCH_3$, $CH_3CHO$, $CH_3CCH$ and $C_3H_6$, it is particularly important to provide better constraints for the rates of these reactions at low temperatures.

**6 Conclusions**

This study reports the results of a combined experimental/theoretical/astrochemical modeling investigation of the C($^3P$) + $CH_3COCH_3$ reaction. Experiments were performed using an existing supersonic flow reactor coupled with pulsed laser photolysis and pulsed laser induced fluorescence for the detection of C($^3P$) and H($^2S$) at vacuum ultraviolet wavelengths to obtain temperature dependent rate constants over the 50-296 K temperature range and temperature dependent branching ratios for H-atom formation over the 75-296 K range. Electronic structure calculations clearly show the barrierless nature of the pathways leading to $^3C_3H_6$ + CO as the major products at all temperatures, with further evolution of $^3C_3H_6$ to ground state $^1C_3H_6$ predicted to occur through intersystem crossing. Several minor channels ($CH_2CCOCH_3$ + H, $CH_3CCH_2$ + H + CO, $C_3H_4$ + HCO + H) could be responsible for the production of a large fraction of the H-atoms observed experimentally. The presence of a weakly bound van der Waals complex between C and $CH_3COCH_3$ might explain the observed temperature dependence of the H-atom yield, through an increased production of H + $HCCHCOCH_3$ as the temperature falls. Further experiments and/or theoretical work will be required to disentangle the relative contributions of these pathways. The effect of the C($^3P$) + $CH_3COCH_3$ reaction on interstellar acetone abundances was tested through a gas-grain dense interstellar cloud model. Using a temperature independent rate constant value of 2.2 × $10^{-10}$ cm$^3$ s$^{-1}$, the simulations showed that the gas-phase acetone abundance decreased by as much as two orders of magnitude at early times, with only a small reduction in the acetone abundance at cloud ages considered to be representative of dense interstellar clouds. Despite its high reactivity, the C($^3P$) + $CH_3COCH_3$ reaction is predicted to be only a minor source of propene in this model.

**Author Information**
**Corresponding Author**
*Email: kevin.hickson@u-bordeaux.fr.

**Supporting Information**
Geometries and frequencies of the stationary points involved in the C($^3P$) + $CH_3COCH_3$ reaction obtained at the M06-2X/aug-cc-pVTZ level of theory (DOCX).

**Acknowledgements**

K. M. H. and V. W. acknowledge support from the French program ''Physique et Chimie du Milieu Interstellaire'' (PCMI) of the CNRS/INSU with the INC/INP co-funded by the CEA and CNES. K. M. H. also acknowledges support from the ''Programme National de Planétologie'' (PNP) of the CNRS/INSU.

Supporting information file for

**Kinetic Study of the Gas-Phase Reaction between Atomic Carbon and Acetone. Low Temperature Rate Constants and Hydrogen Atom Product Yields**

Kevin M. Hickson,[1,*] Jean-Christophe Loison,[1] and Valentine Wakelam[2]

[1]Institut des Sciences Moléculaires ISM, CNRS UMR 5255, Univ. Bordeaux, 351 Cours de la Libération, F-33400, Talence, France

[2]Laboratoire d'astrophysique de Bordeaux, CNRS, Univ. Bordeaux, B18N, allée Geoffroy Saint-Hilaire, F-33615 Pessac, France

**Geometries and frequencies of the stationary points involved in the C($^3$P) + CH$_3$COCH$_3$ reaction obtained at the M06-2X/aug-cc-pVTZ level of theory.**

**Geometries in Cartesian coordinates**

**Reactants and Products**

CH$_3$COCH$_3$
```
 C   0.00000001870384   -0.00000000422835    0.18602837740653
 O  -0.00000014265663   -0.00000006240400    1.39089204428984
 C   0.00000680803691    1.28242233539903   -0.61052890082265
 C  -0.00000677256864   -1.28242231982573   -0.61052891149649
 H   0.00027610415161    2.13692371889576    0.05904464284650
 H  -0.00027597261171   -2.13692366518312    0.05904468196830
 H   0.87688534661262    1.31561307911012   -1.25862162253853
 H  -0.87717918885784    1.31586626815558   -1.25819385241016
 H  -0.87688537049766   -1.31561309983610   -1.25862152655879
 H   0.87717916968751   -1.31586625008319   -1.25819393268455
```

HCCHCOCH$_3$
```
 C  -0.32284520495004    0.10985361037353    0.06496777966095
 O  -0.93160590621430    0.12415554849061    1.10600843444849
 C   0.24427371635435    1.35580942098570   -0.56221636913770
 C  -0.10942151762672   -1.18537455080587   -0.65425071237227
 H   0.01841252901913    2.21263336599183    0.06464848958516
 H   1.32256476810597    1.24893561257046   -0.68282060039485
 H  -0.17915919952443    1.49286535898829   -1.55750866011179
 C   0.52773258330259   -1.27647968549415   -1.79076432049861
 H   0.80630707232789   -2.04754443118176   -2.48959922580050
 H  -0.53077134589444   -2.05715117231864   -0.15624792657888
```

CH$_3$CO

| | | | |
|---|---|---|---|
| C | -0.96838708495203 | -0.66177367442313 | 0.00000078566803 |
| C | 0.00018648349808 | 0.49407271025227 | 0.00000244228037 |
| O | 1.17299516746654 | 0.46561086558913 | -0.00000299443596 |
| H | -0.43472508407789 | -1.61261293010581 | 0.00000070416548 |
| H | -1.60481650590892 | -0.57539675124910 | 0.87856322559353 |
| H | -1.60481297602576 | -0.57539522006337 | -0.87856416327143 |

CCH$_3$

| | | | |
|---|---|---|---|
| C | -2.59449501429966 | 1.68588554347691 | 0.04316231336064 |
| H | -1.50145602031574 | 1.74766547754805 | 0.01793601398080 |
| H | -2.97782061291938 | 2.73154817677500 | 0.03366698248128 |
| H | -3.00024969750658 | 1.19188846828416 | -0.84602630144415 |
| C | -3.20091114115863 | 1.35784550591586 | 1.30477963892142 |

$^3$CH$_3$CCH$_3$

| | | | |
|---|---|---|---|
| C | 0.00000006799431 | -0.00000129000408 | -0.03771929523381 |
| C | 0.00003334436765 | 1.34116324513297 | -0.63234430402035 |
| C | -0.00003338267224 | -1.34116452506699 | -0.63234723065241 |
| H | 0.00009170706693 | 2.11394420578366 | 0.13703158373309 |
| H | -0.00009233909526 | -2.11394717803340 | 0.13702695259266 |
| H | 0.88312222001366 | 1.50708416699662 | -1.26212831539207 |
| H | -0.88308344254480 | 1.50715813912857 | -1.26206947543592 |
| H | -0.88312207699113 | -1.50708373878005 | -1.26213193446062 |
| H | 0.88308358306088 | -1.50715835185730 | -1.26207244013057 |

$^1$CH$_3$CCH$_3$

| | | | |
|---|---|---|---|
| C | -1.04789209463613 | -0.49425235228936 | -0.14870688218515 |
| C | -0.08474595046238 | 0.60655380724189 | -0.27151328873167 |
| C | 1.29735602587700 | 0.13247554650663 | -0.12791913822487 |
| H | 1.44703474603788 | -0.83706148045045 | 0.36574140533310 |
| H | 1.99770279966192 | 0.88469850479069 | 0.22824313613638 |
| H | -0.67043551143556 | -1.51281224672889 | -0.31080688125870 |
| H | -1.26633297154745 | -0.40419947676005 | 0.93170700760220 |
| H | -1.99573561247981 | -0.32482955131073 | -0.65475365714804 |
| H | 1.53616854998453 | -0.01463301439973 | -1.19763413482325 |

CH$_3$CCO

| | | | |
|---|---|---|---|
| C | 0.10042726297591 | 1.16860803964083 | 0.84959068127642 |
| C | -0.08428685218921 | 0.65084016145083 | 1.97652561195830 |
| C | 0.29607935458759 | 1.67443902585970 | -0.49495069410729 |
| H | 0.47647160426826 | 2.74781314472088 | -0.52451028986164 |
| H | 1.13483838574479 | 1.16038571642066 | -0.96952583834220 |
| H | -0.58335795095514 | 1.44900852239294 | -1.10306285507223 |
| O | -0.25638780443220 | 0.18201638951415 | 3.04976738414865 |

CH$_2$CCOCH$_3$

| | | | |
|---|---|---|---|
| C | 0.01111967658826 | 0.09744950414822 | 0.12452014917168 |
| C | -0.49412262597738  |  0.42187972341790  |  1.48809590167114  |
| C |  0.50919913204491  |  1.09836785489686  | -0.59291745960956  |
| C | -0.08861581341910  | -1.33497714775790  | -0.30611835674158  |
| H |  0.53383158982730  |  2.10094746101929  | -0.18825349424557  |
| H |  0.45795802160876  | -1.97563515052388  |  0.38608662047225  |
| H |  0.90013653942045  |  0.93750130084678  | -1.58953104941001  |
| H | -1.12824629778640  | -1.66231109181640  | -0.28994928941910  |
| H |  0.31259169479000  | -1.46699417444436  | -1.30888434984032  |
| O | -0.94552141539680  | -0.32349860808650  |  2.27952922135107  |

CH₃CCH₂

|   |                    |                    |                    |
|---|--------------------|--------------------|--------------------|
| C |  0.43774665947813  |  0.26846974848431  | -1.13401338274001  |
| C |  0.08081970000464  |  1.10571284525637  | -2.06845066440994  |
| C |  0.07588164831151  | -1.07064981196536  | -0.65364283375308  |
| H | -0.28483724171526  | -1.03043483098718  |  0.37443272839324  |
| H | -0.71372356038730  | -1.50413286968462  | -1.27722707526802  |
| H |  0.93584126684096  | -1.74029868744683  | -0.67674177408881  |
| H |  0.54862709958288  |  2.07639157475675  | -2.18344244853136  |
| H | -0.70888557211557  |  0.85485203158656  | -2.77747454960201  |

CH₃

|   |                    |                    |                    |
|---|--------------------|--------------------|--------------------|
| C | -0.70372357606606  |  1.18654111678269  | -0.00005833245105  |
| H | -0.70392912723809  |  0.22215060857474  | -0.47829875439109  |
| H | -0.70309120365291  |  2.08307218215012  | -0.59589698537590  |
| H | -0.70309310774294  |  1.25481377899245  |  1.07425224461803  |

HCO

|   |                    |                    |                    |
|---|--------------------|--------------------|--------------------|
| C | -0.75931040142755  |  0.46707027306888  |  3.74746918555020  |
| O | -0.88701919150079  | -0.60908520484423  |  3.31157666395112  |
| H | -0.22645040707167  |  1.30018493177535  |  3.22146415049867  |

CO

|   |                    |                    |                    |
|---|--------------------|--------------------|--------------------|
| C |  0.00000000000000  |  0.00000000000000  | -0.64095512334366  |
| O |  0.00000000000000  |  0.00000000000000  |  0.48043012334366  |

CH₃CHCH₂

|   |                    |                    |                    |
|---|--------------------|--------------------|--------------------|
| C | -1.13117731438409  | -0.50304606783484  | -0.00000007737906  |
| C |  0.00003244622660  |  0.47546729381995  | -0.00000015594099  |
| C |  1.28305132497653  |  0.14730972848395  |  0.00000001396368  |
| H |  1.59680126178830  | -0.89001897187121  | -0.00000008653904  |
| H |  2.05991253087391  |  0.89934665153453  |  0.00000009567141  |
| H | -0.27130028328953  |  1.52676225069718  |  0.00000020396423  |
| H | -0.76326937018507  | -1.52796761844465  |  0.00000010986960  |
| H | -1.76623667917448  | -0.36388145568398  |  0.87652663560215  |
| H | -1.76623691683216  | -0.36388181070094  | -0.87652673921198  |

H₂CCCH₂

| | | | |
|---|---|---|---|
| C | -0.00000003598378 | -0.00000008390851 | -0.00000000493640 |
| C | -0.00000007095097 | -0.00000001289853 | 1.29929676163271 |
| C | -0.00000000231002 | -0.00000015946924 | -1.29929676527984 |
| H | 0.00000007707970 | 0.92817085726923 | 1.85551889701515 |
| H | 0.00000001275505 | -0.92817080901590 | 1.85551901121975 |
| H | 0.92817084738687 | 0.00000013617379 | -1.85551892747881 |
| H | -0.92817082797685 | 0.00000007184914 | -1.85551897217256 |

CH$_2$CCO
| | | | |
|---|---|---|---|
| C | 0.80855434690393 | -1.64706437466224 | 0.00000001417645 |
| C | -0.00005348679437 | -0.61074604739055 | 0.00000002184009 |
| C | -0.27940664776765 | 0.64581521702713 | 0.00000000086302 |
| O | -0.68659837560138 | 1.73417659841576 | -0.00000001248457 |
| H | 1.88991096433432 | -1.53763544000684 | -0.00000000639553 |
| H | 0.42827719892514 | -2.66369495338325 | -0.00000001799946 |

**Intermediates**

MIN1
| | | | |
|---|---|---|---|
| C | -0.16352211755793 | 0.20987034760863 | -0.31347534523821 |
| C | -0.22108731947873 | 0.60236165019641 | 2.60218793743764 |
| C | 0.16996622468619 | 1.35393460329517 | -1.16985388303839 |
| C | -0.30274581186853 | -1.22610202007185 | -0.57725177251872 |
| H | 0.18803097483591 | 2.27950318802880 | -0.59332640275855 |
| H | -0.56043093802296 | -1.77050076720897 | 0.33243336540257 |
| H | 1.14916836883768 | 1.23753514595904 | -1.64254044966168 |
| H | -0.56802446948755 | 1.48513982200174 | -1.97455538499410 |
| H | -1.08803695346301 | -1.42985576954546 | -1.31559224520469 |
| H | 0.62774948762801 | -1.65888888964936 | -0.96544454600873 |
| O | -0.39939008117693 | -0.47271777547714 | 2.87468376924972 |

MIN2
| | | | |
|---|---|---|---|
| C | -0.01194404223955 | -0.02597801455009 | 0.19465330992908 |
| O | -0.04188676553894 | 0.02640611905240 | 1.39706870818044 |
| C | 0.01671297907638 | 1.23202065357893 | -0.64521316629181 |
| C | 0.00837181977552 | -1.33474868937989 | -0.55368702089927 |
| H | -0.02343448856040 | 2.11624650675674 | -0.01486720753788 |
| H | -0.05868664960317 | -2.16247028855417 | 0.14575591996550 |
| H | 0.94332789297146 | 1.25565346279867 | -1.22998577516096 |
| H | -0.83427123642990 | 1.23783977258277 | -1.33326824392970 |
| H | -0.81550479473865 | -1.37170300786101 | -1.26824927191851 |
| H | 0.92979217075018 | -1.40813154102917 | -1.13499235821083 |
| C | 0.26397582353708 | 0.64443628780483 | -3.30903273692603 |

MIN3
| | | | |
|---|---|---|---|
| C | -0.11065678283129 | 0.20084104570905 | 0.07204906715194 |
| C | -0.19674788130291 | 0.49055199208174 | 1.47955559683294 |
| C | 0.13980670173191 | 1.29770743461363 | -0.89420375911534 |
| C | -0.27245911151351 | -1.19427246281092 | -0.42480906766823 |
| H | 0.26328253228358 | 2.25113644142102 | -0.38679070826301 |
| H | -0.48855232802888 | -1.88493034655268 | 0.38594122745694 |
| H | 1.03405342746663 | 1.08835496395339 | -1.48797438814062 |
| H | -0.68907078041443 | 1.37726906024182 | -1.60383156389423 |
| H | -1.07766023163798 | -1.24093266328813 | -1.16295358105459 |
| H | 0.63813720678993 | -1.51859557990479 | -0.93702996547719 |
| O | -0.39083527774303 | -0.25219401236414 | 2.38925994087140 |

MIN4
| | | | |
|---|---|---|---|
| C | 0.12059702647454 | -0.06664226190554 | 0.01713671986137 |
| O | 0.66907895736751 | 0.11194041280523 | 1.07178484200812 |
| C | -0.32182936744251 | 1.11833137827412 | -0.84711006255168 |
| C | -0.12944035086533 | -1.43602786417605 | -0.55694765327389 |
| H | -0.38186105839683 | 1.99949265258287 | -0.20185223750340 |
| H | 0.36396468250534 | -2.19236255336504 | 0.04587404087216 |
| H | 1.70002298869454 | 1.54878802777759 | -1.92272815132838 |
| H | -1.31335825964576 | 0.91963717733416 | -1.26255910876161 |
| H | -1.20469948784583 | -1.62392633990840 | -0.57629843515892 |
| H | 0.22474134906780 | -1.47480213037206 | -1.58775497806642 |
| C | 0.64923622908652 | 1.30514276215312 | -1.93136281889733 |

MIN 5
| | | | |
|---|---|---|---|
| C | -0.15657516278513 | 0.46738146260138 | 0.51822439119640 |
| C | 0.02617229290588 | 0.86784709033301 | 1.88501828989984 |
| C | -0.02134511638168 | 1.16825987432879 | -0.77903011957709 |
| H | -0.12077152005574 | 2.24868230694659 | -0.63360607947005 |
| H | 0.95445648352060 | 0.96322208578218 | -1.21404985786900 |
| H | -0.79122692607216 | 0.83548951140384 | -1.47363032822508 |
| O | -1.24691909481922 | 0.66812698493920 | 1.35851572059409 |
| C | 0.50404549748821 | 0.16607807352891 | 3.09888852152559 |
| H | 1.56049707264839 | 0.37409976425595 | 3.25610293104664 |
| H | -0.05072595943572 | 0.49574275425308 | 3.97622027957707 |
| H | 0.37357398068656 | -0.91476987087293 | 2.98448958550159 |

MIN6
| | | | |
|---|---|---|---|
| C | -0.12463087188148 | 0.15893614946121 | 0.03938980393740 |
| C | -0.19338575786233 | 0.52661731439984 | 1.42913457298081 |
| C | 0.11831461633300 | 1.17680714857325 | -0.97427097100752 |
| C | -0.29244153014339 | -1.26507253472189 | -0.36195727801087 |
| H | -0.51421459749441 | -1.88170777654750 | 0.50533793004828 |
| H | -1.09280268111813 | -1.35974457669740 | -1.09966532522754 |
| H | 0.61853533017613 | -1.62978596814996 | -0.84457991940526 |
| O | -0.38255954998991 | -0.27202769712582 | 2.33646225833097 |
| H | -0.70243955113243 | 1.66461271153213 | -1.48246915070644 |
| H | 1.12288208681322 | 1.40040012270234 | -1.30746214376535 |

S46

| H | -0.05856836670029 | 1.59904673697382 | 1.64401615492550 |

MIN7
| C | -0.54364038222218 | 0.80548793677536 | 0.88061628083906 |
| O | -1.73634667571567 | 0.53793031881531 | 0.98399923502504 |
| C | 0.14441224373359 | 1.79620911500529 | 1.78268202877433 |
| H | 0.58828023508339 | 2.59626056207417 | 1.18947957651043 |
| H | 0.95296972189795 | 1.30479567212607 | 2.32469735092881 |
| H | -0.57332542931216 | 2.21084258635973 | 2.48482128166017 |
| C | 0.23942819919762 | 0.15971478089893 | -0.12812321508159 |
| C | -0.09900576590784 | -0.83444473347858 | -1.14550506302486 |
| H | -0.22107215699801 | -0.36586251375677 | -2.12573067659317 |
| H | 0.68219614853484 | -1.59012254026429 | -1.23785803288996 |
| H | -1.04115795999155 | -1.32205255665525 | -0.88202106694829 |

MIN8
| C | -0.08245107525222 | 0.09718581310793 | 0.21262805746180 |
| C | 0.15556939368973 | 0.22893019611552 | 1.64072941120433 |
| C | -0.02847987738273 | 1.29708002069527 | -0.68095854588172 |
| C | -0.02580943144878 | -1.24573312012321 | -0.44629829017121 |
| H | -0.17910371561867 | 2.20677756050973 | -0.10681936762173 |
| H | -0.17464417576616 | -2.03490998807569 | 0.28504241871462 |
| H | 0.94640811668560 | 1.34104493827751 | -1.16816505974155 |
| H | -0.79006585082403 | 1.22225470553272 | -1.45890777893067 |
| H | -0.78771865733780 | -1.31628101630602 | -1.22436600785401 |
| H | 0.94907402111123 | -1.37639133983094 | -0.91765369845970 |
| O | -1.13348127305616 | 0.19497810319719 | 1.29398165997984 |

MIN9
| C | -0.13494793441388 | 0.11643782715097 | 0.23828590867773 |
| C | 0.34268871008592 | -0.50101345339544 | 2.46730298713615 |
| C | -0.09727839069314 | 1.32227079371883 | -0.61382128980674 |
| C | -0.01913466719471 | -1.28521054802228 | -0.20894926366721 |
| H | -0.55173804526651 | 2.16774340314182 | -0.09885798866409 |
| H | -0.68859959844050 | -1.93777523378295 | 0.35468497107859 |
| H | 0.93092336089497 | 1.60519727297025 | -0.87059572144875 |
| H | -0.63290485077226 | 1.14391278966461 | -1.54439958777963 |
| H | -0.25341800048811 | -1.35102672168265 | -1.26872542576091 |
| H | 0.99399121199514 | -1.66981611735746 | -0.04312424714455 |
| O | 0.11041920429307 | 0.38927898759431 | 1.58819965737942 |

MIN10
| C | 0.01083177769884 | -0.09398772883272 | 0.19921141952008 |
| O | -0.24055923127547 | 0.03752761779081 | 1.38858212693245 |
| C | 0.23959083055543 | 1.07406444022281 | -0.62992149532193 |
| C | 0.10114249561133 | -1.45450492316359 | -0.44983506153444 |
| H | 0.16276866883957 | 2.02467025359794 | -0.11021158093282 |
| H | -0.09941950411689 | -2.22148509661805 | 0.29214198140943 |
| H | 1.58571895264801 | 1.02723715215505 | -2.38209893856354 |
| H | -0.61500655780464 | -1.53072773320592 | -1.26932371853483 |
| H | 1.09365270646054 | -1.60254442255108 | -0.87834809934347 |
| C | 0.55818047052610 | 1.02951799791307 | -2.04306509507652 |
| H | -0.21398060914283 | 1.12738244269167 | -2.79453153855441 |

**Transition states**

TS1

| | | | |
|---|---|---|---|
| C | -0.02157759685125 | 0.15416441172135 | -0.18624044746157 |
| C | -0.27484824473130 | 0.61209851551254 | 2.01809300507047 |
| C | 0.18011166386765 | 1.34540123564966 | -1.01585112532492 |
| C | -0.25297033948131 | -1.25866325901748 | -0.50422352372878 |
| H | 0.29884354172314 | 2.23140773542231 | -0.39185686405879 |
| H | -0.37235091646954 | -1.84948369421169 | 0.40529545680152 |
| H | 1.07366040313958 | 1.25619933532410 | -1.64471010055112 |
| H | -0.67060643749137 | 1.51996466559346 | -1.68558790191707 |
| H | -1.15859981754281 | -1.39242811675229 | -1.10771394811489 |
| H | 0.58245844526398 | -1.68722891124239 | -1.07006853752488 |
| O | -0.53482322662675 | -0.31649604489956 | 2.61207678551005 |

TS2

| | | | |
|---|---|---|---|
| C | 0.27262608519621 | -0.28210094987248 | 0.11355480680924 |
| O | 0.66255141723526 | -0.05409084897067 | 1.22940256438585 |
| C | 0.04801399893804 | 0.83107549636850 | -0.88585851350720 |
| C | -0.00437921828531 | -1.69364011086630 | -0.34686076467446 |
| H | -0.67024484287591 | 1.53568744949690 | -0.46816767144953 |
| H | 0.15843076335015 | -2.40176787408041 | 0.46080382064166 |
| H | 0.98646602239344 | 1.36627518160999 | -1.02643923215456 |
| H | -0.31193973849281 | 0.46747929000778 | -1.84626933586882 |
| H | -1.03875434038494 | -1.76895603072388 | -0.69841585590709 |
| H | 0.66518091895966 | -1.93489091387362 | -1.18010756510521 |
| C | -0.76796372403378 | -1.95102254649570 | -2.99087661786977 |

TS3

| | | | |
|---|---|---|---|
| C | -0.01216756818671 | -0.00057101884850 | 0.10486339588755 |
| C | -0.02402684122938 | 0.63179224530685 | 1.38792015913411 |
| C | 0.24084250956662 | 1.18843535068251 | -0.74323424121513 |
| C | -0.19212741755148 | -1.42048661447564 | -0.26008155656999 |
| H | -0.40378559744557 | -2.02923476636087 | 0.61660830765342 |
| H | -1.00739471244639 | -1.52925401859614 | -0.97956987477955 |
| H | 0.70831048619003 | -1.80225494082355 | -0.74929176708862 |
| O | -0.16624115489786 | 0.32436142764931 | 2.53226410645961 |
| H | -0.57818901674412 | 1.55763652267879 | -1.35712041947901 |
| H | 1.22037124145208 | 1.28645811550847 | -1.20643730809714 |
| H | 0.21440807129278 | 1.79311769727878 | 0.65407919809475 |

TS4

| | | | |
|---|---|---|---|
| C | 0.16563369185183 | 0.04209239904451 | 0.39075589142075 |
| C | 0.40146613254568 | 0.19671602770864 | 1.78873918182217 |
| C | 0.15236792559749 | 1.23958925902647 | -0.49711257592622 |
| C | -0.04376476899877 | -1.30121532115649 | -0.22213071055923 |
| H | -0.58716483653565 | 1.94392771848778 | -0.11431222139455 |
| H | -0.07432246240199 | -2.07877010922965 | 0.53419463057264 |
| H | 1.12408791990313 | 1.73318730321860 | -0.44279013534460 |
| H | -0.08242506607034 | 0.99225164074656 | -1.53087395182665 |
| H | -0.95981775327640 | -1.31496964683094 | -0.81693322154914 |
| H | 0.78911877323246 | -1.49262263804576 | -0.90483209417500 |
| O | -0.88517955584746 | 0.03981336703027 | 1.81529520695983 |

TS5

| | | | |
|---|---|---|---|
| C | -0.57723961983756 | 0.25597434047492 | 0.02162043946092 |
| C | -0.30363336686632 | -0.46158003066576 | 1.51885275244909 |
| C | 0.15414940613164 | 1.42474628000417 | -0.51896294624965 |
| H | 0.68186798595266 | 1.93939628019393 | 0.28798793726087 |
| H | 0.86940896835157 | 1.10927908747411 | -1.27590097916773 |
| H | -0.55559324388937 | 2.11721941019613 | -0.97257875347054 |
| O | -1.41618149315962 | 0.30861986652782 | 1.03572035052099 |
| C | 0.18330645844411 | -1.44139373544678 | 0.11187264900726 |
| H | 0.96855374157916 | -1.93131786725359 | 0.69106491708572 |
| H | -0.65010296637757 | -2.12144304575898 | -0.04412863225107 |
| H | 0.64546412967131 | -1.19950058574596 | -0.85554773464586 |

TS6

| | | | |
|---|---|---|---|
| C | 0.37930005978385 | -0.67321093906050 | -0.13192750876799 |
| C | -0.20490492922462 | 0.49587562961606 | 0.34527784676188 |
| C | -0.01748791108349 | -1.70025597309374 | -1.10766404084329 |
| H | -0.96284783873277 | -1.42940602418063 | -1.58606032868157 |
| H | 0.76258459957178 | -1.79481474051078 | -1.86385312726609 |
| H | -0.12655635364395 | -2.66720684538076 | -0.61599316984841 |
| O | -1.06805895446058 | -0.37866835182483 | 0.84942826433979 |
| C | 0.22834466271520 | 1.78660172600969 | 0.91427004913341 |
| H | 0.75360731685972 | 2.37892309997907 | 0.16842479853088 |
| H | -0.64247727309253 | 2.34378722857347 | 1.25767255778346 |
| H | 0.89849662130739 | 1.63837518987294 | 1.77042465885793 |

TS7

| | | | |
|---|---|---|---|
| C | -0.16920368622223 | 0.65023005175528 | 1.15617291562314 |
| C | -0.26573174974063 | -1.61687498669426 | -0.18426036627984 |
| O | 0.64855022859571 | 0.55866893344910 | 2.05312399896373 |
| C | 0.15784527859469 | 1.46231961037544 | -0.42307796776448 |
| H | -0.73173940753021 | 2.00409662095873 | -0.72260664930097 |
| H | 0.63005624635221 | 0.96754202861226 | -1.27444105909187 |
| H | 0.89633511470446 | 2.12893894827580 | 0.02434417254563 |
| H | -0.08863773751584 | -1.70774221549377 | -1.25556717830842 |
| H | -1.09808054754365 | -2.27146610052982 | 0.07904369283376 |
| H | 0.61948609793627 | -1.95024355163457 | 0.36209102395589 |
| C | -0.59887983763077 | -0.22546933907419 | 0.18517741682344 |

TS8
| | | | |
|---|---|---|---|
| C | -1.15757098986948 | 0.01618966361030 | 1.33013514538973 |
| O | -1.79173802275774 | 0.23982010983573 | 2.29813445048962 |
| C | 0.10268602577897 | 1.87232504537468 | 1.08365203271864 |
| H | -0.56158596494367 | 2.40340854120599 | 0.41685652163180 |
| H | 1.06937080114373 | 1.59049354639355 | 0.69360648843823 |
| H | 0.03700653882049 | 2.10874957816886 | 2.13417465748673 |
| C | -0.77999929776105 | -0.70238013862638 | 0.28897412689358 |
| C | 0.16997316166137 | -0.58551996068424 | -0.81750229626533 |
| H | 1.19318555264065 | -0.66810680078548 | -0.43726331467693 |
| H | 0.02300468240991 | -1.36674432525546 | -1.55956776169207 |
| H | 0.08840569117681 | 0.39052336866239 | -1.30414235121409 |

TS9
| | | | |
|---|---|---|---|
| C | 0.02448350751026 | 0.42574800620320 | 0.56970565997971 |
| C | -0.13543811668171 | 0.47004080282159 | 1.85288889208921 |
| C | 0.23268436427164 | 1.31894917782047 | -0.57416841807246 |
| C | -0.13881174254197 | -1.64140041235171 | -0.47447500647789 |
| H | 0.61845482847078 | 2.29794690709183 | -0.29628518431813 |
| H | -0.43987821965822 | -2.27302090354659 | 0.34667793634788 |
| H | 0.88660890823638 | 0.84933238811940 | -1.30843829125514 |
| H | -0.73225386346809 | 1.45852267506146 | -1.07247665673276 |
| H | -0.88868257386416 | -1.40401522595671 | -1.21563755696398 |
| H | 0.88060364023965 | -1.73605214178374 | -0.81967771390382 |
| O | -0.30777073251457 | 0.23394872652080 | 2.99188633930739 |

TS10
| | | | |
|---|---|---|---|
| C | -0.29987839300943 | -0.17887124491863 | 0.48194239270653 |
| O | -0.30880537180875 | -1.13442607288596 | -0.11295043913999 |
| H | 0.60868376481817 | 1.31329731780460 | -0.36899195356653 |

TS11
| | | | |
|---|---|---|---|
| C | -0.08890570239885 | 0.04563111784284 | 0.28587182751047 |
| C | 0.81648136000391 | -0.00270414482523 | 2.14389665853944 |
| C | 0.06063000820920 | 1.28003824785828 | -0.50427865090724 |
| C | -0.15832929712199 | -1.33042581774495 | -0.25618233911434 |
| H | -0.87202004222350 | 1.85333969849011 | -0.50100467201856 |
| H | -0.30821653028241 | -2.04903708251015 | 0.54488693590739 |
| H | 0.83030390772269 | 1.91021675461725 | -0.05014069504099 |
| H | 0.33607591103565 | 1.05743484826600 | -1.53247521467542 |
| H | -0.98533209523965 | -1.41812360116061 | -0.96953224597858 |
| H | 0.75612735140519 | -1.58840697042606 | -0.79600101562097 |
| O | -0.38681487111025 | 0.24203694959255 | 1.63495941139881 |

TS12

| | | | |
|---|---|---|---|
| C | -0.28273125451093 | 0.08581218035497 | 0.11529694825741 |
| C | -0.00725209477758 | -0.48336037466888 | 2.55186465119477 |
| C | -0.02763589187501 | 1.33018568403370 | -0.62015866744981 |
| C | 0.01167194329562 | -1.29876707839591 | -0.26040974856800 |
| H | -0.39683410272052 | 2.18732664134082 | -0.05819856567741 |
| H | -0.08741761614116 | -1.94653699580795 | 0.61495633377662 |
| H | 1.04472222228447 | 1.47680911158530 | -0.79802155129218 |
| H | -0.52373659012351 | 1.31201792942258 | -1.59204604659397 |
| H | -0.67519922586983 | -1.65307985501928 | -1.03092483386637 |
| H | 1.03177316551626 | -1.39930629499819 | -0.64849847645202 |
| O | -0.08736055507780 | 0.38889905215282 | 1.72613995667097 |

TS13

| | | | |
|---|---|---|---|
| C | -0.27974749649867 | 0.08527152187477 | 0.11097796096649 |
| C | -0.06472185822333 | -0.48253751629921 | 2.55071863787987 |
| C | -0.01488039228403 | 1.32995064553898 | -0.62041352007761 |
| C | 0.02768505576230 | -1.29808490236967 | -0.25978058717574 |
| H | -0.30945035069812 | 2.19159893054588 | -0.02220893211646 |
| H | 0.00195177965410 | -1.93142298759167 | 0.63206922760636 |
| H | 1.04923265812210 | 1.43084334332358 | -0.86589751460223 |
| H | -0.57182698554192 | 1.35400352021768 | -1.55932542444652 |
| H | -0.69688648260559 | -1.68936483257813 | -0.97674889140901 |
| H | 1.02185713372948 | -1.37725069984080 | -0.71288856948575 |
| O | -0.16321206141632 | 0.38699197717862 | 1.72349761286063 |

TS14

| | | | |
|---|---|---|---|
| C | 0.21789079861456 | 0.17145943073301 | -1.30331086537127 |
| C | 0.08545693314143 | 1.17607051169649 | -2.11247277934457 |
| C | 0.28772134641381 | -0.88630279945121 | -0.53004471874948 |
| H | -0.25162204877023 | -0.91977418917644 | 0.40729996701364 |
| H | -1.04529363914429 | -2.14075338486883 | -1.21791495879958 |
| H | 1.02754363206515 | -1.65415262254911 | -0.71392538143456 |
| H | 0.68908458603392 | 2.06885505265471 | -2.00826631673073 |
| H | -0.63931160835436 | 1.14450800096138 | -2.91792494658342 |

**Frequencies in cm$^{-1}$**

**Reactants and products**

| CH$_3$COCH$_3$ | HCCHCOCH$_3$ | CH$_3$CO | CCH$_3$ | $^3$CH$_3$CCH$_3$ | $^1$CH$_3$CCH$_3$ | CH$_3$CCO | CH$_2$CCOCH$_3$ | CH$_3$CCH$_2$ | CH$_3$ | HCO | CO | CH$_3$CHCH$_2$ | H$_2$CCCH$_2$ | CH$_2$CCO |
|---|---|---|---|---|---|---|---|---|---|---|---|---|---|---|
| 94 | 114 | 110 | 509 | 157 | 245 | 34 | 151 | 178 | 432 | 1106 | 2271 | 193 | 370 | 120 |
| 166 | 164 | 469 | 896 | 159 | 268 | 168 | 188 | 321 | 1412 | 1994 | | 430 | 370 | 256 |
| 386 | 273 | 870 | 1114 | 298 | 445 | 503 | 241 | 513 | 1413 | 2724 | | 590 | 867 | 508 |
| 496 | 495 | 961 | 1314 | 855 | 677 | 530 | 404 | 905 | 3144 | | | 938 | 871 | 696 |
| 540 | 499 | 1049 | 1324 | 938 | 703 | 789 | 457 | 920 | 3321 | | | 947 | 871 | 960 |
| 808 | 541 | 1355 | 1448 | 950 | 952 | 848 | 596 | 942 | 3323 | | | 959 | 1014 | 1051 |
| 893 | 723 | 1459 | 2913 | 1056 | 1009 | 1064 | 754 | 1055 | | | | 1036 | 1014 | 1073 |
| 901 | 745 | 1461 | 3020 | 1126 | 1155 | 1391 | 803 | 1109 | | | | 1080 | 1122 | 1487 |
| 1089 | 858 | 1993 | 3065 | 1315 | 1208 | 1467 | 965 | 1389 | | | | 1193 | 1420 | 1824 |
| 1127 | 911 | 3075 | | 1395 | 1325 | 1473 | 1008 | 1424 | | | | 1332 | 1484 | 2293 |
| 1251 | 1022 | 3171 | | 1402 | 1342 | 1552 | 1031 | 1461 | | | | 1406 | 2082 | 3111 |
| 1391 | 1053 | 3173 | | 1460 | 1396 | 2214 | 1071 | 1476 | | | | 1455 | 3152 | 3188 |
| 1399 | 1224 | | | 1463 | 1401 | 3059 | 1268 | 1764 | | | | 1482 | 3156 | |
| 1464 | 1262 | | | 1471 | 1493 | 3105 | 1407 | 3037 | | | | 1495 | 3233 | |
| 1471 | 1394 | | | 1474 | 1500 | 3139 | 1439 | 3095 | | | | 1739 | 3233 | |
| 1475 | 1466 | | | 2982 | 2972 | | 1477 | 3115 | | | | 3066 | | |
| 1495 | 1475 | | | 2988 | 2977 | | 1491 | 3138 | | | | 3120 | | |
| 1846 | 1680 | | | 3026 | 3032 | | 1717 | 3202 | | | | 3150 | | |
| 3060 | 1823 | | | 3028 | 3037 | | 1951 | | | | | 3170 | | |
| 3063 | 3067 | | | 3102 | 3149 | | 3081 | | | | | 3179 | | |
| 3121 | 3116 | | | 3104 | 3150 | | 3144 | | | | | 3258 | | |
| 3127 | 3132 | | | | | | 3169 | | | | | | | |
| 3176 | 3181 | | | | | | 3186 | | | | | | | |
| 3177 | 3272 | | | | | | 3278 | | | | | | | |

**Intermediates**

| MIN1 | MIN2 | MIN3 | MIN4 | MIN5 | MIN6 | MIN7 | MIN8 | MIN9 | MIN10 |
|---|---|---|---|---|---|---|---|---|---|
| 53 | 87 | 92 | 67 | 144 | 129 | 79 | 168 | 128 | 126 |
| 74 | 114 | 139 | 91 | 162 | 198 | 137 | 216 | 136 | 167 |
| 86 | 152 | 247 | 236 | 233 | 268 | 186 | 362 | 200 | 255 |
| 93 | 201 | 259 | 329 | 251 | 304 | 194 | 375 | 260 | 382 |
| 136 | 240 | 355 | 392 | 426 | 357 | 400 | 426 | 309 | 425 |
| 165 | 390 | 419 | 523 | 449 | 411 | 574 | 449 | 368 | 513 |
| 175 | 499 | 589 | 592 | 788 | 498 | 586 | 629 | 553 | 552 |
| 317 | 543 | 789 | 788 | 799 | 615 | 769 | 793 | 699 | 577 |
| 855 | 806 | 944 | 800 | 972 | 831 | 957 | 949 | 927 | 706 |
| 945 | 885 | 961 | 922 | 1032 | 900 | 982 | 993 | 957 | 817 |
| 958 | 900 | 1010 | 1027 | 1081 | 963 | 1041 | 1014 | 984 | 949 |
| 1060 | 1081 | 1046 | 1068 | 1091 | 1002 | 1062 | 1094 | 1043 | 997 |
| 1131 | 1127 | 1290 | 1105 | 1169 | 1043 | 1140 | 1232 | 1169 | 1039 |
| 1318 | 1245 | 1312 | 1175 | 1189 | 1284 | 1368 | 1259 | 1329 | 1120 |
| 1398 | 1372 | 1396 | 1249 | 1392 | 1299 | 1383 | 1407 | 1367 | 1272 |
| 1406 | 1395 | 1409 | 1270 | 1398 | 1386 | 1422 | 1411 | 1418 | 1353 |
| 1461 | 1466 | 1466 | 1387 | 1474 | 1427 | 1453 | 1432 | 1420 | 1401 |
| 1466 | 1471 | 1469 | 1447 | 1476 | 1449 | 1464 | 1472 | 1458 | 1447 |
| 1474 | 1484 | 1486 | 1465 | 1478 | 1478 | 1471 | 1475 | 1473 | 1466 |
| 1478 | 1495 | 1498 | 1475 | 1479 | 1480 | 1480 | 1495 | 1486 | 1478 |
| 2242 | 1853 | 1844 | 1852 | 1583 | 1635 | 1645 | 1509 | 1491 | 1675 |
| 3004 | 3037 | 3047 | 3054 | 3047 | 3000 | 3046 | 3055 | 3021 | 3077 |
| 3009 | 3075 | 3051 | 3075 | 3048 | 3055 | 3081 | 3059 | 3040 | 3137 |
| 3047 | 3083 | 3092 | 3104 | 3132 | 3103 | 3107 | 3120 | 3102 | 3152 |
| 3050 | 3135 | 3099 | 3140 | 3134 | 3154 | 3129 | 3125 | 3116 | 3156 |
| 3120 | 3180 | 3169 | 3193 | 3168 | 3178 | 3144 | 3168 | 3163 | 3195 |
| 3125 | 3193 | 3177 | 3233 | 3168 | 3249 | 3190 | 3169 | 3171 | 3243 |

**Transition states**

| TS1 | TS2 | TS3 | TS4 | TS5 | TS6 | TS7 | TS8 | TS9 | TS10 | TS11 | TS12 | TS13 | TS14 |
|---|---|---|---|---|---|---|---|---|---|---|---|---|---|
| -246 | -75 | -2061 | -1046 | -596 | -2162 | -864 | -663 | -529 | -609 | -1992 | -681 | -675 | -672 |
| 16 | 74 | 107 | 166 | 174 | 99 | 37 | 89 | 66 | 355 | 29 | 90 | 95 | 258 |
| 105 | 85 | 188 | 222 | 226 | 159 | 197 | 169 | 121 | 2214 | 147 | 112 | 139 | 361 |
| 121 | 148 | 236 | 331 | 266 | 221 | 204 | 197 | 132 | | 164 | 134 | 178 | 409 |
| 129 | 253 | 246 | 368 | 329 | 261 | 217 | 220 | 191 | | 230 | 248 | 253 | 446 |
| 147 | 390 | 447 | 415 | 404 | 401 | 388 | 294 | 228 | | 346 | 285 | 287 | 892 |
| 239 | 499 | 530 | 469 | 528 | 427 | 522 | 397 | 402 | | 442 | 309 | 318 | 896 |
| 314 | 532 | 669 | 794 | 646 | 782 | 583 | 430 | 460 | | 767 | 529 | 531 | 931 |
| 844 | 799 | 797 | 947 | 788 | 947 | 871 | 477 | 480 | | 921 | 848 | 852 | 1000 |
| 947 | 886 | 866 | 993 | 937 | 980 | 876 | 581 | 540 | | 949 | 953 | 958 | 1023 |
| 948 | 932 | 908 | 1033 | 985 | 1011 | 908 | 794 | 746 | | 989 | 958 | 961 | 1101 |
| 1045 | 1074 | 971 | 1057 | 997 | 1058 | 1016 | 861 | 824 | | 1023 | 1053 | 1051 | 1417 |
| 1126 | 1118 | 1037 | 1234 | 1043 | 1106 | 1064 | 930 | 986 | | 1032 | 1126 | 1127 | 1475 |
| 1319 | 1240 | 1118 | 1256 | 1120 | 1299 | 1276 | 1051 | 1051 | | 1305 | 1333 | 1335 | 2034 |
| 1395 | 1366 | 1281 | 1386 | 1333 | 1379 | 1360 | 1378 | 1389 | | 1319 | 1398 | 1404 | 3138 |
| 1400 | 1387 | 1324 | 1406 | 1391 | 1395 | 1393 | 1423 | 1413 | | 1405 | 1419 | 1423 | 3153 |
| 1459 | 1473 | 1398 | 1439 | 1424 | 1450 | 1439 | 1426 | 1430 | | 1415 | 1459 | 1461 | 3217 |
| 1463 | 1478 | 1425 | 1470 | 1458 | 1466 | 1462 | 1429 | 1443 | | 1463 | 1466 | 1467 | 3235 |
| 1471 | 1482 | 1470 | 1480 | 1474 | 1468 | 1465 | 1467 | 1467 | | 1465 | 1473 | 1474 | |
| 1475 | 1487 | 1482 | 1489 | 1477 | 1484 | 1475 | 1487 | 1498 | | 1483 | 1476 | 1478 | |
| 2147 | 1853 | 1756 | 1499 | 1531 | 1552 | 1716 | 1970 | 2097 | | 1488 | 1545 | 1544 | |
| 3009 | 3032 | 1840 | 3061 | 2994 | 3016 | 3042 | 3042 | 3045 | | 3027 | 3015 | 3018 | |
| 3014 | 3084 | 3054 | 3068 | 3061 | 3049 | 3052 | 3095 | 3113 | | 3032 | 3017 | 3020 | |
| 3053 | 3086 | 3101 | 3116 | 3067 | 3125 | 3109 | 3122 | 3117 | | 3081 | 3083 | 3087 | |
| 3056 | 3159 | 3104 | 3140 | 3133 | 3129 | 3130 | 3158 | 3153 | | 3087 | 3097 | 3096 | |
| 3118 | 3171 | 3162 | 3161 | 3163 | 3143 | 3143 | 3283 | 3274 | | 3163 | 3112 | 3109 | |
| 3127 | 3183 | 3183 | 3196 | 3175 | 3170 | 3210 | 3295 | 3287 | | 3173 | 3143 | 3144 | |